\documentclass[prx,twocolumn,aps,showpacs,amsmath,amssymb]{revtex4}

\usepackage{bm}
\usepackage{graphicx}
\usepackage{amsmath}
\usepackage{times}

\usepackage{amsmath, amssymb,mathtools}
\usepackage[colorlinks,bookmarks=false,citecolor=blue,linkcolor=red,urlcolor=blue]{hyperref}
\usepackage[usenames,dvipsnames,svgnames,table]{xcolor}

\newcommand\underrel[2]{\mathrel{\mathop{#2}\limits_{#1}}}

\begin{document}

\title{Quantum variance: a measure of quantum coherence and quantum correlations for many-body systems}

\author{Ir\'en\'ee Fr\'erot$^1$ and Tommaso Roscilde$^{1,2}$}
\affiliation{$^1$ Laboratoire de Physique, CNRS UMR 5672, Ecole Normale Sup\'erieure de Lyon, Universit\'e de Lyon, 46 All\'ee d'Italie, 
Lyon, F-69364, France}
\affiliation{$^2$ Institut Universitaire de France, 103 boulevard Saint-Michel, 75005 Paris, France}

\begin{abstract}
Quantum coherence is a fundamental common trait of quantum phenomena, from the interference of matter waves to quantum degeneracy of identical particles. Despite its importance, estimating and measuring quantum coherence in generic, mixed many-body quantum states remains a formidable challenge, with fundamental implications in areas as broad as quantum condensed matter, quantum information, quantum metrology and quantum biology. Here we provide a quantitative definition of the variance of quantum coherent fluctuations (the \emph{quantum variance}) of any observable on generic quantum states. The quantum variance generalizes the concept of thermal de Broglie wavelength (for the position of a free quantum particle) to the space of eigenvalues of any observable, quantifying the degree of coherent delocalization in that space. The quantum variance is generically measurable and computable as the difference between the static fluctuations and the static susceptibility of the observable; despite its simplicity, it is found to provide a tight lower bound to most widely accepted estimators of ``quantumness" of observables (both as a feature as well as a resource), such as the Wigner-Yanase skew information and the quantum Fisher information.    
When considering bipartite fluctuations in an extended quantum system, the quantum variance expresses genuine quantum correlations (of discord type) among the two parts. In the case of many-body systems it is found to obey an area law at finite temperature, extending therefore area laws of entanglement and quantum fluctuations of pure states to the mixed-state context. Hence the quantum variance paves the way to the measurement of macroscopic quantum coherence and quantum correlations in most complex quantum systems.   
\end{abstract}


\maketitle

 \section{Introduction}
 Quantum mechanics introduces two fundamentally new traits in a physical system: 1) an intrinsic uncertainty on the knowledge of observables (Heisenberg's uncertainty or coherent quantum fluctuations), and 2) a superior form of correlation among degrees of freedom, stemming from correlated quantum uncertainties (or entanglement) \cite{Feynmanbook,Schroedinger1935,Scaranibook}. Quantum uncertainty of observables persists even at zero temperature in the form of so-called zero-point fluctuations, responsible for 
 macroscopic quantum phenomena such as the inability of liquid Helium to solidify at ambient pressure \cite{Leggettbook}. On the other hand two quantum systems (hereafter called $A$ and $B$) can exhibit correlations far exceeding any classical counterpart, which for pure quantum states are embodied by entanglement \cite{Horodecki2009}.
 The supremacy of both fluctuations and correlations in quantum systems, as compared to classical ones, is at the heart of the complexity of many-body quantum states, challenging all realms of quantum physics, from relativistic quantum field theory to atomic/molecular physics and quantum condensed matter. At the same time quantum fluctuations and correlations (going beyond entanglement \cite{Modietal2012}) are by now recognized as essential ingredients for the supremacy of quantum devices over classical ones, in the context of both quantum information processing \cite{Modietal2012} and quantum metrology \cite{PezzeS2014}. 
 
  Despite their fundamental as well as practical importance, quantum coherence and quantum correlations remain very hard to both quantify theoretically and to measure experimentally. Quantum uncertainty of an observable and quantum entanglement between two subsystems are generically defined only for pure states \cite{Feynmanbook,Horodecki2009}. In the case of generic, real-life mixed states, the most widespread concept of quantum coherence is related to the thermal de Broglie wavelength (TdBWL) \cite{Huangbook}, expressing the spatial extent of coherent quantum fluctuations for a single quantum particle in free space; but this concept does not even extend to systems as simple as a particle in a potential. More recently, several definitions of mixed-state quantum coherence have been put forward \cite{WignerY1963,Luo2003,Girolamietal2013,Girolami2014,LeviM2014,Baumgratzetal2014}, which nonetheless share the generically prohibitive requirement of knowing the full density matrix of the state, and they are therefore limited to few-body systems. As for the entanglement of mixed states, one can only provide sufficient conditions (witnesses) for the presence of entanglement between the components of the system \cite{Peres1996,VidalW2002,GuehneT2009}. Yet, even for non-entangled mixed states it has been recognized that quantum correlations may exist, associated with the violation of classical information-theory identities, and quantified via the so-called quantum discord \cite{Modietal2012} and discord-like quantities \cite{Girolamietal2013,Girolamietal2014}.  Despite their deep conceptual meaning, entanglement and quantum correlations of mixed states remain in general information-theoretical objects, generically accessible (to calculations and measurements alike) only when defined between two (or a few) elementary quantum degrees of freedom
\cite{Wootters1998,Fukuharaetal2015,Islametal2015,Sarandyetal2013,Gessneretal2014,Girolamietal2014}. 

 \begin{center}
  \begin{figure*}
  \includegraphics[width=18cm]{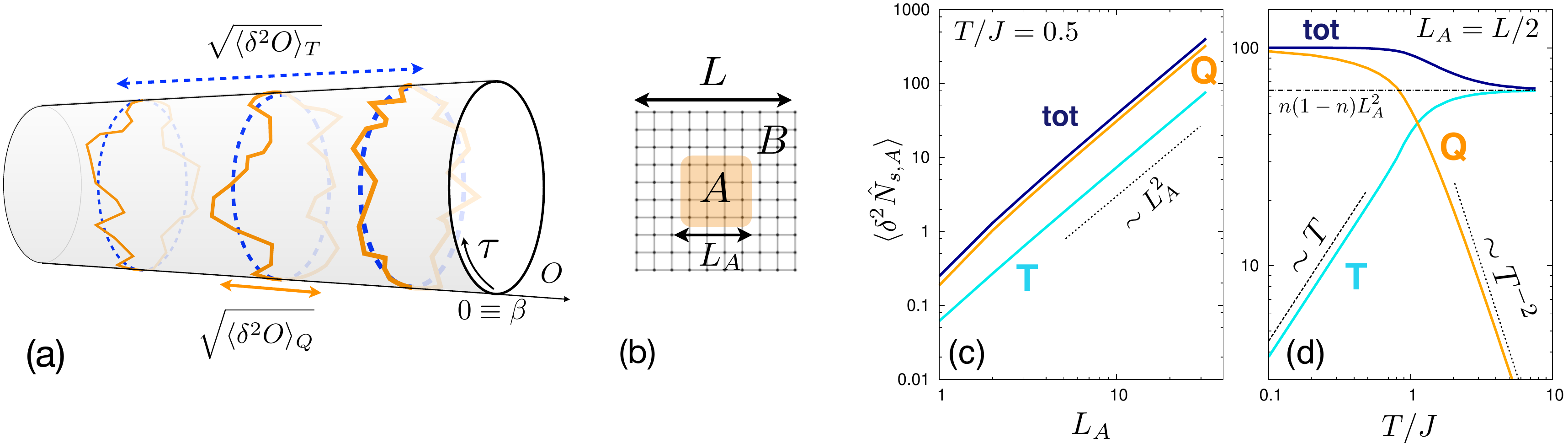}
  \caption{{\bf Thermal vs. quantum fluctuations.} (a) Different imaginary-time paths $O(\tau)$ in the space of eigenvalues of the observable $\hat{O}$ are shown, associated with the path-integral representation of a generic mixed state $\hat{\rho}$. While the thermal/incoherent fluctuations $\langle \delta^2 O \rangle_T$ are associated with the fluctuations of the centroids of the paths (dashed blue lines), the quantum/coherent fluctuations $\langle \delta^2 O \rangle_Q$ are associated with the fluctuations of the paths around their centroids; (b) Geometry of the $A$-$B$ bipartition of an extended quantum system used in the text;  (c) Scaling of the total (tot), thermal (T) and quantum (Q) fluctuations of the staggered particle number $N_{s,A}$ on a subsystem $A$ of size $L_A$ for hardcore bosons on a square lattice at temperature $T/J=0.5$ (the system is defined on a $L\times L$ torus with $L=32$). All fluctuation terms exhibit volume-law scaling. Here and in the following graphs, the error bar is smaller than or comparable to the line thickness; (d) Temperature dependence of the $N_{s,A}$ fluctuations for a subsystem of linear size $L_A = L/2$. The dashed line indicates the infinite temperature limit, in which each lattice site fluctuates independently, with a shot-noise variance $n(1-n)$ where $n=1/2$ is the lattice filling.}
  \label{f.fig1}
  \end{figure*}
  \end{center}
  
 Here we show that both quantum coherent fluctuations and quantum correlations in generic quantum states can be quantified in terms of elementary physical concepts. The variance of fluctuations in generic mixed states possesses in fact an \emph{additive} structure, in which the incoherent/thermal part can be separated from the coherent/quantum part, which we name \emph{quantum variance} (QV). The QV is defined in terms of the violation of a classical, static fluctuation-dissipation relation, and as such it is fully computable and measurable for generic systems.  The QV of a given observable is a measure of its genuine quantum uncertainty in mixed states, and, in the case of bipartite fluctuations, it represents a measure of correlated quantum uncertainties, namely of quantum correlations. Remarkably the QV is convex { (namely it \emph{decreases} upon incoherent mixing of states with the same QV)}, and it gives a tight lower bound to both the Wigner-Yanase skew information \cite{WignerY1963,Luo2003} and to the quantum Fisher information \cite{BraunsteinC1994} which are widely accepted, yet generically prohibitive measures (for both calculations and experiments) of the quantumness of observables and of correlations \cite{Chen2005,Toth2012,Hyllusetal2012,Girolamietal2013,Girolami2014,Girolamietal2014}. 
 
 The structure of the paper is as follows.
 Sec.~\ref{s.separation} introduces the separation between coherent and incoherent fluctuations, and the definition of QV;
 Sec.~\ref{s.properties} reviews the fundamental properties of the QV as a lower bound to known measures of ``quantumness" of observables;
 Sec.~\ref{s.volume} describes the volume-law scaling of QV for generic extensive observables;
 Sec.~\ref{s.critical} illustrates the fundamental separation of scales between thermal and quantum fluctuations at a thermal critical point;
 Sec.~\ref{s.area} discusses the area-law scaling of the QV of bipartite fluctuations;
Sec.~\ref{s.correlations} illustrates the link between the QV and other measures of quantum correlations in the case of free fermions;  and finally Sec.~\ref{s.conclusions} elaborates on the general link between QV on the one side, and entanglement and quantum correlations on the other; and on the experimental measurement of the QV with cold-atom quantum simulators. The technical aspects are kept to a minimum level in the main text, and they are postponed to the Appendices. 
 
  \section{Separating classical and quantum fluctuations.} 
  \label{s.separation}
  Let us first show that, given a density matrix $\hat\rho$ such that $\langle ... \rangle = {\rm Tr} [\hat{\rho} (...)]/ {\rm Tr}\hat\rho$, and a generic Hermitian operator $\hat{O}$, the fluctuations of the latter can be written as
  \begin{equation}
  \langle \delta^2 \hat{O} \rangle = \langle \hat{O}^2 \rangle - \langle \hat{O}\rangle^2 = \langle \delta^2 \hat{O} \rangle_T + \langle \delta^2 \hat{O} \rangle_Q~
  \end{equation}
  where $\langle \delta^2 \hat{O}  \rangle_T$ represents thermal/incoherent fluctuations, while $\langle \delta^2 \hat{O} \rangle_Q$ represent quantum/coherent fluctuations. In the following we shall focus our attention on thermal equilibrium states, but the whole treatment is readily generalizable to arbitrary density matrices (see App.~\ref{a.pathintegral}). If $\hat{\rho} = e^{-\beta \hat{\cal H}}/{\cal Z}$ (${\cal Z} = {\rm Tr}(e^{-\beta \hat{\cal H}})$) is the thermal density matrix of a system of Hamiltonian $\hat{\cal H}$ at temperature $k_B T = 1/\beta$, and $[\hat{O},\hat{\cal H}]=0$, it is well known that the fluctuations of $\hat{O}$ satisfy a (classical) fluctuation-dissipation theorem
  \begin{equation}
  \langle \delta^2 \hat O \rangle = \chi_O~ k_B T
    \label{e.fluctdiss}
  \end{equation}
 where $\chi_O = \partial^2 F / \partial h^2 |_{h=0}$ is the susceptibility associated with the application of a term $-h \hat O$ to the Hamiltonian, and $F = -k_B T \log {\cal Z}$ is the free energy.  On the other hand, if $[\hat{O},\hat{\cal H}]\neq 0$ the quantum uncertainty on the value of $\hat{O}$ adds up to the thermal fluctuations, and, as a result
  \begin{equation}
  \langle \delta^2 O \rangle \geq  \chi_O~ k_B T = \frac{1}{\beta} \int_0^{\beta} d\tau~ \langle \delta \hat{O}(\tau) \delta \hat{O}(0) \rangle =: \langle \delta^2 \hat{O} \rangle_T
  \label{e.thermal}
  \end{equation}
where $\hat{O}(\tau) = e^{\tau \hat{\cal H}} \hat{O} e^{-\tau \hat{\cal H}}$ is the operator evolved in imaginary time. Eq.~\eqref{e.thermal} shows that thermal fluctuations do not exhaust the total fluctuations of the observable. It is then natural to define the QV for the observable $\hat{O}$ as the residual fluctuations, or as the violation of the classical fluctuation-dissipation relation of Eq.~\eqref{e.fluctdiss}:  
\begin{equation}
\langle \delta^2 \hat{O} \rangle_Q  
=  \langle \delta^2 \hat{O} \rangle -   \chi_O~ k_B T~.
\label{e.qvariance}
\end{equation}
The QV has a particularly suggestive interpretation in the context of a path-integral representation of the partition function of the system, using a basis of Hilbert space which diagonalizes the $\hat{O}$ operator (see Fig.~\ref{f.fig1}(a)). As discussed in App.~\ref{a.pathintegral}, this allows one to cast the partition function in the form: 
\begin{equation}
{\cal Z} = \int {\cal D}[O(\tau)] e^{-S[O(\tau), \partial_\tau O(\tau), ...]}
\end{equation}
where $O(\tau)$ is a periodic trajectory ($O(0) = O(\beta)$) in the space of eigenvalues of ${\hat O}$, associated with the Feynman path in the basis diagonalizing $\hat O$, and $S$ is the associated action weighting the trajectory. When assigning a path-integral expression to each of the terms in Eq.~\eqref{e.qvariance}, one can easily find that (see App.~\ref{a.pathintegral})
\begin{equation}
\langle \delta^2 \hat{O} \rangle_Q = \left\langle \frac{1}{\beta} \int d\tau \left( O(\tau) -   {\bar O}\right)^2 \right \rangle_S 
\label{e.qvariance2}
\end{equation}
where $\langle ... \rangle_S$ is the average over the ensemble of paths $O(\tau)$ weighed by the action $S$, and
\begin{equation}
\bar O = \bar O [O(\tau)] = \frac{1}{\beta} \int d\tau ~O(\tau)
\end{equation}
is the centroid of the path \cite{FeynmanHibbs}. 
Eq.~\eqref{e.qvariance2} shows that the QV represents the (squared) amplitude of the imaginary-time fluctuations of the trajectory $O(\tau)$ around the path centroid. Clearly such fluctuations have a genuine quantum origin \cite{FeynmanHibbs, CaoV1994, Cuccolietal1995}. If $\hat{O}$ is the position $\hat{x}$ of a one-dimensional particle, in App.~\ref{a.deBroglie}  we show that $\langle \delta^2 \hat{x} \rangle_Q \sim \lambda_{\rm dB}^2$, namely the QV is tightly related to the quantum uncertainty on the position expressed by the TdBWL $\lambda_{\rm dB}$. When moving to higher dimensions and generic quantum systems, the QV generalizes therefore the concept of TdBWL (or quantum coherence length) to the space of eigenvalues of any Hermitian operator. And, most remarkably, it does so in a computable and measurable manner, being expressed as the difference between a fluctuation property and a response function. 

\section{Properties of the quantum variance.} 
\label{s.properties}

The QV represents a physically measurable lower bound to fundamental quantities in quantum information. The Dyson-Wigner-Yanase skew information \cite{Wehrl1978}
\begin{equation}
I_{\alpha}(\hat{O},\hat{\rho}) = - \frac{1}{2} {\rm Tr} \{ [\hat{O},\hat{\rho}^{\alpha}][\hat{O},\hat{\rho}^{1-\alpha}] \}
\end{equation}
 with $\alpha\in [0,1]$, generalizing the Wigner-Yanase skew information ($\alpha=1/2$) \cite{WignerY1963}, probes the quantum uncertainty of $\hat{O}$ stemming from its non-commutativity with $\hat{\rho}$. As shown in App.~\ref{a.skewinfo}, the QV is simply 
 \begin{equation}
 \langle \delta^2 \hat{O} \rangle_Q[\hat\rho] = \int_0^1 d\alpha ~ I_{\alpha}~.
 \end{equation}
  From the convexity of $I_{\alpha}$ \cite{Wehrl1978} follows the convexity of the QV. Moreover one can prove that 
  \begin{equation}
  \langle \delta^2 \hat{O} \rangle_Q[\hat\rho] \leq  I_{1/2}(\hat{O},\hat{\rho})
  \end{equation}
  (the equality holding for pure states). Finally, the quantum Fisher information \cite{BraunsteinC1994}
  \begin{equation}
  F_Q(\hat O;\hat \rho) = 2 \sum_{nm} |\langle n | \hat{O} |m\rangle|^2 (p_n-p_m)^2/(p_n+p_m)
 \end{equation} 
 (where  $\hat\rho = \sum_n p_n |n\rangle \langle n|$)  expresses the sensitivity of the density matrix to a unitary transformation $\hat{U}(h) = e^{-ih\hat O}$ generated by the observable $\hat O$, and it quantifies the fundamental metrological gain in using the state $\hat{\rho}$ to estimate the parameter $h$ \cite{PezzeS2014}. As shown in App.~\ref{a.QFI},
 \begin{equation}
 \langle \delta^2 \hat{O} \rangle_Q[\hat\rho] \leq F_Q(\hat O;\hat \rho)/4~.
 \end{equation} 
 The inequality becomes an equality for pure states. As discussed later, the inequalities satisfied by the QV have considerable implications concerning its importance for entanglement witnessing and metrological applications. Conversely, the computability and measurability of QV gives unprecedented insight into the skew and quantum Fisher information for quantum many-body systems.  

 \section{Quantum variance of a global observable and volume law.} 
 \label{s.volume}
 
 Due to its inherent quantum nature, the QV exhibits very special size and temperature dependences. In the following we shall concentrate on thermal equilibrium states, and we start our analysis with the case of a generic, macroscopic observable ${\hat O}$ that does \emph{not} commute with the Hamiltonian of the system. As an example we consider the case of two-dimensional hardcore bosons on the square lattice:
 \begin{equation}
 \hat {\cal H} = -J  \sum_{\langle ij \rangle} \left( \hat{b}_i^{\dagger} \hat{b}_j + {\rm h. c.} \right)
 \label{e.hcb}
 \end{equation} 
 where $\hat{b}_i$, $\hat{b}^{\dagger}_i$ are hardcore boson operators, satisfying the (anti)-commutation relations $\{ b_i, b_i^{\dagger} \} = 1$ and $[b_i, b_j] = [b_i,b_j^{\dagger}] = 0$ ($i\neq j$). We treat this model with a numerically exact quantum Monte Carlo algorithm based on the Stochastic Series Expansion approach \cite{SyljuasenS2002}, which allows us to investigate the imaginary-time dynamics of many-body observables \cite{Sandvik1992}. The Hamiltonian $\hat{\cal H}$ does not commute with any finite-wavevector Fourier component of the density profile, and in particular with the staggered particle number 
 \begin{equation}
 \hat{N}_{\rm s} = \sum_i (-1)^i \hat{b}^{\dagger}_i \hat{b}_i.
 \end{equation} 
 To investigate the scaling of fluctuations (both thermal and quantum) we isolate a subsystem $A$ of linear size $L_A$ in a larger system (of linear size $L$ -- see Fig.~\ref{f.fig1}(b)), and we investigate the scaling of local observables/fluctuations in $A$ with the size of the $A$ region itself. This approach allows one to extract scaling properties while using a single simulation box, and it is also directly applicable to experiments giving access to local properties, such as those based on quantum-gas microscopy \cite{Bakretal2009}.
 
  Fig.~\ref{f.fig1}(c) shows that both the thermal and the quantum contribution to fluctuations obey a \emph{volume-law} scaling in the example at hand:
  \begin{equation}
  \langle \delta^2 \hat{N}_{s,A} \rangle_T, ~\langle \delta^2 \hat{N}_{s,A} \rangle_Q \sim L_A^d
  \end{equation}
  where $\hat{N}_{\rm s,A} = \sum_{i\in A} (-1)^i \hat{b}^{\dagger}_i \hat{b}_i$.
  A volume-law scaling of quantum fluctuations is generically expected when the observable of interest is extensive, and its Heisenberg's uncertainty is the result of the non-commutativity between an extensive set of terms in the Hamiltonian and in the observable in question.  The separation between thermal and quantum fluctuations gives rise to a very meaningful result when tracking the temperature dependence of the fluctuations on a subsystem of fixed size ($L_A = L/2$ in this case). As shown in Fig.~\ref{f.fig1}(c), the thermal component grows linearly with $T$ at low $T$, while the quantum component decreases monotonically with $T$ starting from the zero-point fluctuations. Most remarkably, in the example at hand quantum fluctuations are found to dominate the total fluctuations, and they lead to a \emph{monotonically} decreasing behavior of $\langle \delta^2 \hat{N}_{\rm s} \rangle $, in complete contradiction with the classical expectation that fluctuations should grow with temperature at low $T$. 

\begin{center}
  \begin{figure}
  \includegraphics[width=\columnwidth]{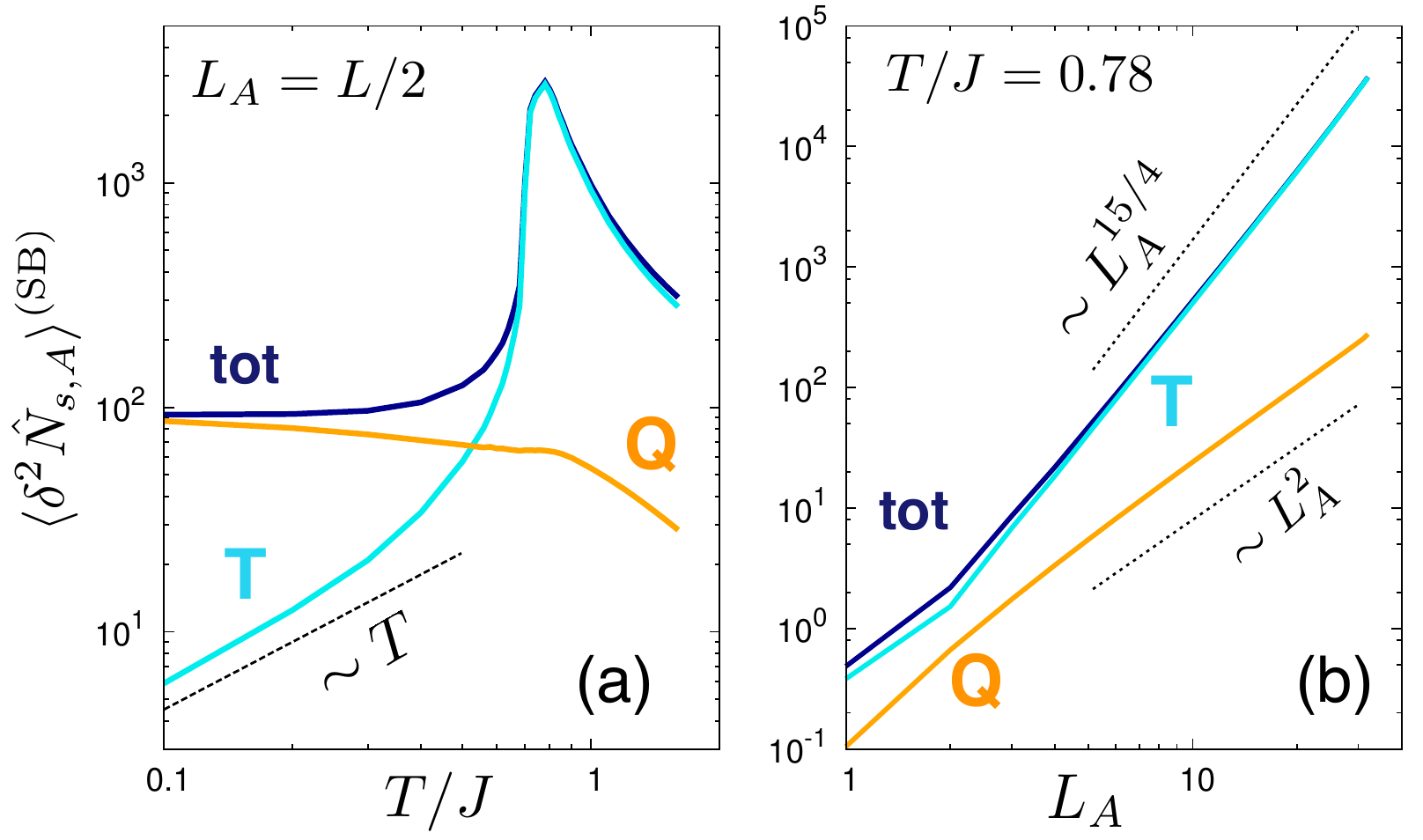}
  \caption{{\bf Critical thermal fluctuations, non-critical quantum fluctuations.} (a) Temperature dependence of the order parameter fluctuations at the Ising transition of 2$d$ hardcore bosons with nearest neighbor repulsion $V=2.1 J$; the sharp peak marks the transition at $T_c \approx 0.78 J$; (b) Scaling of fluctuations close to the critical temperature: the total and thermal fluctuations are found to scale as $L_A^{d+\gamma/\nu}$ with 
  $\gamma = 7/4$ and $\nu=1$ for the 2$d$ Ising universality class.}
  \label{f.Ising}
  \end{figure}
  \end{center}

 \section{Quantum variance does not go critical at a thermal transition.} 
\label{s.critical}

Having shown that QV generically obeys a volume law for extensive non-conserved observables, one can naturally ask what is the fate of QV at a thermal critical point, at which thermal fluctuations of the order parameter become \emph{super}-extensive. If the QV only captures the quantum mechanical part of fluctuations of the order parameter, one would naturally expect that its scaling is \emph{not} modified at a thermal transition, given that the latter is purely driven by thermal fluctuations. To answer to this question, we consider a quantum many-body model exhibiting a thermal phase transition with an order parameter not commuting with the Hamiltonian; this is readily obtained by generalizing the hardcore-boson Hamiltonian of Eq.~\eqref{e.hcb} to include a nearest-neighbor repulsion $V$:
\begin{equation}
 \hat {\cal H} = -J  \sum_{\langle ij \rangle} \left( \hat{b}_i^{\dagger} \hat{b}_j + {\rm h. c.} \right) + V  \sum_{\langle ij \rangle} \left( \hat{n}_i-\frac{1}{2} \right ) \left ( \hat{n}_j - \frac{1}{2} \right ) ~.
 \end{equation} 
When $V> 2J$ the model has an Ising phase transition at finite temperature, marking the onset of a checkerboard density wave, with an order parameter given by the staggered density $\hat{N}_s$. Hence, as in the previous section, it is meaningful to investigate the temperature and size scaling of the fluctuations of the local staggered density $\hat{N}_{s}$. In particular, to mimic the behavior in the thermodynamic limit (in which $\langle \hat{N}_{s,A} \rangle \neq 0$), we focus on the fluctuations around a finite-size estimate of the order parameter in the symmetry-breaking (SB) phase, given by 
$\langle | \hat{N}_{s,A} | \rangle$:
\begin{equation}
\langle \delta^2 N_{s,A} \rangle^{\rm(SB)} = \langle \hat{N}^2_{s,A} \rangle - \langle | \hat{N}_{s,A} | \rangle^2~.
\end{equation}
Fig.~\ref{f.Ising}(a) shows that the total and thermal fluctuations of the order parameter exhibit a sharp peak at the Ising transition temperature, while the QV is very smooth at the transition. { In particular, the QV is a monotonically decreasing function of temperature, dramatically showing that quantum fluctuations \emph{cannot} grow under incoherent thermal mixing of the density matrix, even when such fluctuations become singular at a critical point. The monotonicity with temperature is in fact a different condition -- required by physical arguments -- than the convexity discussed in Sec~\ref{s.properties}, which only accounts for the evolution under linear mixing.}  A closeup on the scaling close to the critical point (Fig.~\ref{f.Ising}(b)) finds that the total and thermal fluctuations exhibit the critical super-extensive scaling $L_A^{d+\gamma/\nu}$, where $\gamma$ and $\nu$ are the critical exponents for the susceptibility and correlation length. On the other hand the quantum variance maintains a  volume-law scaling as in the non-critical regime. Therefore a critical point marks a net separation of scales between thermal and quantum fluctuations of the order parameter, the latter being essentially irrelevant in the thermodynamic limit. This observation substantiates the common wisdom that quantum mechanics is irrelevant for the universal properties at thermal critical points, and it shows that order parameters close to a critical point have the nature of emergent classical observables.

\begin{center}
  \begin{figure*}
  \includegraphics[width=18cm]{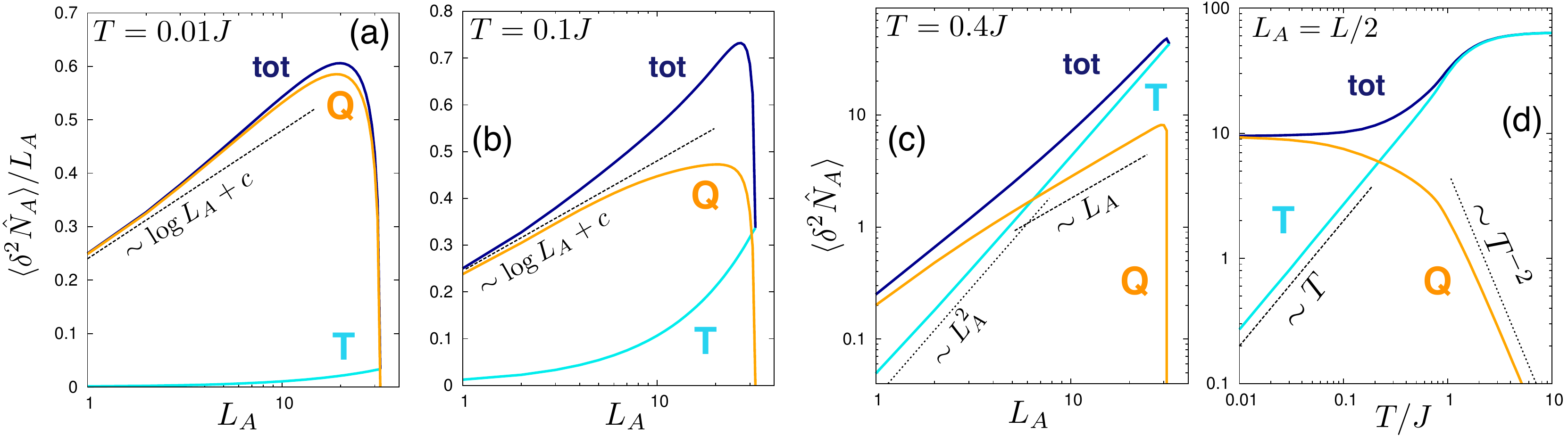}
  \caption{{\bf Bipartite fluctuations.} (a)-(c) Scaling of local particle-number fluctuations in a subsystem $A$ for square-lattice hardcore bosons ($V=0$) at three different temperatures ($T/J = 0.01$, 0.1 and 0.4). Other parameters as in Fig.~\ref{f.fig1}(c); (d) Temperature dependence of the particle-number fluctuations; same parameters as in  Fig.~\ref{f.fig1}(d).}
  \label{f.N}
  \end{figure*}
  \end{center}

\section{Quantum variance of bipartite fluctuations and area law.}  
\label{s.area}

The scaling properties of the QV change drastically when considering the case of bipartite fluctuations of an otherwise globally conserved quantity, such as the particle number $\hat{N}$. Such fluctuations have been the subject of several recent studies in view of their relationship to entanglement in the case of pure states \cite{Songetal2010, Songetal2012,Racheletal2012,FrerotR2015}, as well as at finite temperature, for which a suggestion of how to extract the quantum contribution to fluctuations has been proposed in Ref.~\cite{Songetal2012}.  
In the case of mixed states, $[\hat{N},\hat{\cal H}]=0$ implies automatically that $\langle \delta^2 \hat{N} \rangle_Q=0$. Taking then \emph{any} bipartition of the system into $A$ and $B$ subsystems, imaginary-time fluctuations of the local particle numbers $N_A$  and $N_B$ are perfectly anticorrelated, so that the QV in each subsystem is the same, $\langle \delta^2 \hat{N}_A \rangle_Q = \langle \delta^2 \hat{N}_B \rangle_Q$. Perfect correlation in the quantum uncertainties of $N_A$ and $N_B$ suggests that the QV captures genuine quantum correlations between $A$ and $B$ whenever applied to bipartite fluctuations of globally conserved quantities. Remarkably Fig.~\ref{f.N}(a-c) shows that the QV of bipartite fluctuations scales like the \emph{boundary} between $A$ and $B$, thereby obeying a so-called area law 
\begin{equation}
  \langle \delta^2 \hat{N}_{A} \rangle_Q \sim L_A^{d-1}~,
  \end{equation}
 namely the extensive (volume-law) part of bipartite fluctuations is entirely of incoherent origin. This strongly suggests that the QV captures the fluctuations associated with coherent particle exchanges at the boundary between $A$ and $B$. For the hardcore-boson problem at hand, such fluctuations obey a logarithmically corrected area law at $T=0$ (when all fluctuations are quantum) \cite{Racheletal2012,FrerotR2015},
 \begin{equation}
  \langle \delta^2 \hat{N}_{A} \rangle_Q \sim L_A^{d-1} \log L_A
  \end{equation}
  turning then into an area-law scaling at finite $T$. Nonetheless a logarithmic violation can still be observed at sufficiently low temperature and for small sizes of the $A$ region -- namely smaller than the thermal correlation length $\xi$ for density fluctuations \footnote{Indeed for hardcore bosons the density-density correlation function at finite temperature exhibits a finite correlation length, $\langle \delta n_i \delta n_j \rangle \sim \exp(-|{\bm r}_i-{\bm r}_j|/\xi)$, even though the phase correlations exhibit a critical behavior below the Kosterlitz-Thouless temperature. This was already noticed in Ref.~\cite{Cuccolietal2003}.}.   
 Interestingly, the area-law scaling of the QV (either straight or logarithmically violated) is found to dominate the scaling of total fluctuations at sufficiently small sizes $L_A$ of the subsystem $A$, as shown in Fig.~\ref{f.N}(a-c). This makes the (logarithmically violated) area law of bipartite quantum fluctuations observable under experimentally realistic conditions. 
     
\begin{center}
  \begin{figure}
  \includegraphics[width=\columnwidth]{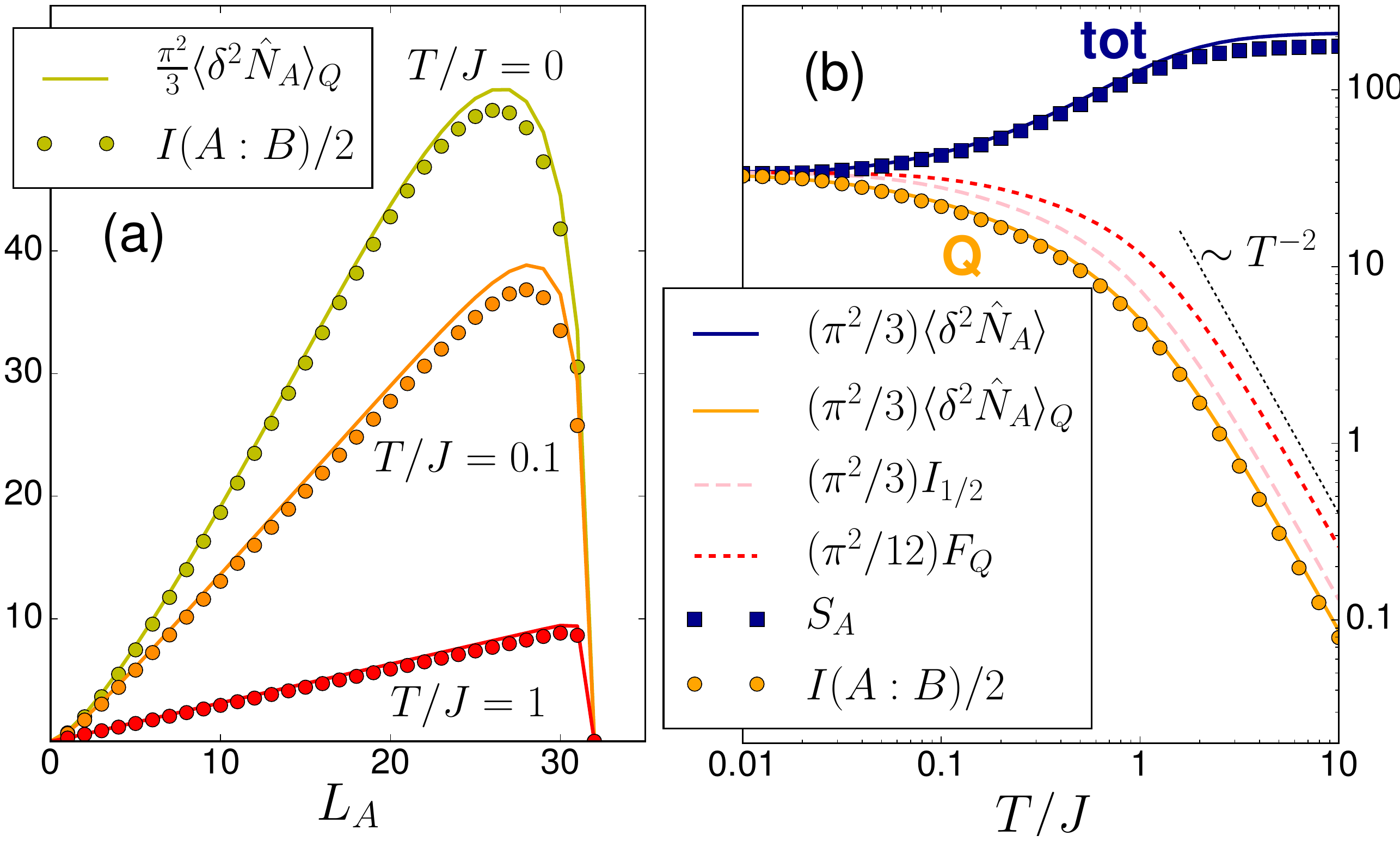}
  \caption{{\bf Quantum correlations vs. quantum mutual information}. (a) Scaling of the quantum variance of bipartite particle-number fluctuations and of the quantum mutual information in a system of free fermions on a $L\times L$ square lattice ($L=32$) at half filling for three different temperatures ($J$ is the hopping amplitude); (b) temperature dependence of the same two quantities, along with the total entropy $S_A$, the total fluctuations $\langle \delta^2 \hat{N}_A \rangle$, the Wigner-Yanase skew information $I_{1/2}(\hat{N}_A,\hat\rho)$, and the quantum Fisher information $F_Q(\hat{N}_A;\hat \rho)$, for an $A$ region with linear size $L_A = L/2$. The $T^{-2}$ decay of the mutual information at high temperature has been proven rigorously for free fermions in Ref.~\cite{Bernigauetal2015}, and it is proven for the quantum variance, skew information and quantum Fisher information in App.~\ref{a.bipartitefluctu}.}
  \label{f.mutual_info}
  \end{figure}
  \end{center}

\section{Quantum variance provides quantum correlations.}          
\label{s.correlations}
 The area-law scaling of bipartite QV of mixed states suggests a link to the similar scaling exhibited by entanglement entropy in ground states of local Hamiltonians \cite{Eisertetal2010}. Yet measures of entanglement at finite temperature (such as the negativity \cite{VidalW2002}) do not admit a simple physical interpretation in terms of entropy of quantum fluctuations (but see below for further discussion on QV and entanglement). As already pointed out, QV is rather connected to quantum correlations, a more general concept than entanglement. The existence of such correlations is captured by the quantum discord \cite{Modietal2012}, given by the difference between the quantum mutual information, $I(A:B) = S_A + S_B - S_{AB}$ (or the entropy sub-extensivity due to correlations between $A$ and $B$) and the classical mutual information $J(A:B)$ (or the maximum amount of information gained on $A$ by performing measurements on $B$). Here $S_{A(B)} = -{\rm Tr} \hat{\rho}_{A(B)} \log \hat{\rho}_{A(B)} $ is the entropy of the reduced density matrix of subsystem $A(B)$, and $S_{AB}$ is the total entropy. The operation of maximization implicit in the definition of quantum discord makes it generically non-computable when $A$ and $B$ are extended subsystems of a quantum many-body system. 

On the other hand, in some special systems the existence of quantum correlations is witnessed by more accessible quantities. Indeed we argue that, in the case of an \emph{ideal} gas, any form of correlation stems from the quantum statistics, while it is trivially absent in the classical limit. The existence of correlations between $A$ and $B$ is generically captured by the quantum mutual information, whose nonzero value is then a direct proof of quantum correlations existing in the system. \footnote{Even though the quantum discord of an ideal gas is not equal to the mutual information \cite{Durganandini2012}, the vanishing of quantum discord in the classical limit is the trivial result of both $I(A:B)$ and $J(A:B)$ being vanishing quantities. Hence the existence of a finite $I(A:B)$ and $J(A:B)$ is already a proof of quantum correlations, and the further aspect that they are not identical -- leading to quantum discord -- is a generic feature of quantum systems.} In the case of an ideal lattice gas the existence of correlations between $A$ and $B$ stems physically from the coherent exchange of particles at the $A$-$B$ boundary, and hence it is tightly linked to the quantum fluctuations of particle numbers. In the following we shall particularly focus on the case of a free Fermi gas on a lattice at half filling, for which the mutual information and QV of particle-number fluctuations can be easily calculated via exact diagonalization \cite{PeschelE2009}. 

The quantum mutual information of free fermions exhibits an area law at finite temperature, as recently proven in Ref.~\cite{Bernigauetal2015, Leschkeetal2015}. Fig.~\ref{f.mutual_info} shows that the area law of mutual information and of QV of particle-number fluctuations are related, as the prefactors of the thermal area laws, $I(A:B)/2 \approx a_I(T) L_A$ and $\langle \delta^2\hat{N}_A \rangle_Q \approx a_N(T) L_A$, are proportional at all $T$, $a_I \approx \frac{\pi^2}{3} a_N$. Remarkably, this is the same relationship holding between the particle-number variance and the entanglement entropy for free fermions at $T=0$ (Fig.~\ref{f.mutual_info}(a)) \cite{Songetal2012, Calabreseetal2012}, and between total entropy and variance in a degenerate Fermi gas (Fig.~\ref{f.mutual_info}(b)). Hence the particle-number QV provides experimental access to the mutual information of free fermions at finite $T$, as much as the total variance of particle-number fluctuations gives access to the entanglement entropy in the ground state. 
Moreover, as shown in Fig.~\ref{f.mutual_info}(b) the QV provides a meaningful lower bound to both the Wigner-Yanase skew information and to the quantum Fisher information; in particular at high temperatures we find that 
$\langle \delta^2\hat{N}_A \rangle_Q \approx \frac{2}{3} I_{1/2}(\hat{N}_A,\hat{\rho})$ and $\langle \delta^2\hat{N}_A \rangle_Q \approx \frac{1}{3} (F_Q(\hat{N}_A;\hat \rho)/4)~$.

\section{Conclusions and outlook.} 
\label{s.conclusions}
In conclusion, we have introduced the quantum variance of generic observables, generalizing the concept of quantum Heisenberg uncertainty to the case of mixed states -- and acting as the ``thermal de Broglie wavelength" in the space of eigenvalues of arbitrary observables. In the case of bipartite fluctuations, the QV expresses the quantum correlations among the two subsystems in arbitrary mixed states. The quantum uncertainty may dominate the fluctuations in quantum many-body systems, leading to a completely non-classical behavior (fluctuations decreasing with temperature, scaling of fluctuations obeying area laws or logarithmically violated area laws, etc.).  

Owing to its definition in terms of fully measurable quantities (fluctuations and response function, see Eq.~\eqref{e.qvariance}), the QV is readily accessible to state-of-the-art experiments. 
All the requirements for the measurement of the QV, and in particular of its scaling in a bipartite setting, are met by trapped-ion experiments \cite{Monroeetal2014} as well as quantum-gas microscope experiments \cite{Bakretal2009}, enabling access to local degrees of freedom. As an example, in microscopy experiments recent progress \cite{Preissetal2015} has demonstrated the ability to resolve different single-site occupation numbers ($n=0,.., 3$) in an optical lattice, providing access to local fluctuations. Moreover the local response function can be probed by an increase of the local uniform (or staggered) chemical potential in a given region $A$ of the system, making use of holographic masks \cite{Bakretal2009} (see App.~\ref{a.exp} for further details).
Hence the total and thermal fluctuations are independently accessible, as well as the scaling of their difference (the QV) with subsystem size (Figs.~\ref{f.fig1}(c-d), \ref{f.N} and \ref{f.mutual_info}). 

 At the theory level, the QV represents a most accessible way to assess quantum correlations, entanglement, and the metrological use of quantum many-body states. As proposed in Refs.~\cite{Girolamietal2013,Girolamietal2014}, the minimal skew information and quantum Fisher information associated with local observables in a subsystem $A$ are both discord-type measures of quantum correlations, and the latter dictates the minimal precision on the estimation of the parameter of an arbitrary local unitary operation; the QV offers a natural measurable lower bound to both quantities (see App.~\ref{a.skewinfo} and \ref{a.QFI} for further discussion). Moreover both the skew information \cite{Chen2005} and the quantum Fisher information \cite{Hyllusetal2012,Toth2012} of collective spin variables witness entanglement among $k$ qubits when exceeding a $k$-dependent bound: a similar violation of the bound by the QV is therefore an even stronger witness -- see App.~\ref{a.enta} for a detailed discussion. 
 
 The QV lends itself to analytical as well as to large-scale numerical simulations based \emph{e.g.} on quantum Monte Carlo - as shown in the present work. Hence its study can be readily extended to generic quantum many-body systems at equilibrium, including interacting fermions, quantum spin models, etc. as well as to non-equilibrium mixed states. While we have mostly focused our attention on bipartite correlations, an extension of our study to multipartite correlations can also be readily achieved by introducing the concept of quantum covariance, as we will further develop in future work. This opens the perspective of a complete characterization of quantum correlations in extended quantum systems, based on experimentally accessible quantities.    
 
 \section{Acknowledgements.} 
 
 TR is indebted to A. Cuccoli, V. Tognetti, R. Vaia and P. Verrucchi for introducing him to seminal ideas on quantum fluctuations and path integrals, and to P. Hauke and A. Sanpera for useful discussions. This work is supported by ANR (``ArtiQ" project). All numerical calculations have been performed at the PSMN center (ENS Lyon).

\appendix

\section{Operator approach to coherent vs. incoherent fluctuations in mixed states: area law of quantum coherence}
\label{a.coh}

As pointed out in the main text, in the case of mixed states described by a density matrix $\hat\rho$ there is a fundamental distinction between thermal/incoherent fluctuations and quantum/coherent fluctuations of any observable which does not commute with $\hat\rho$. This distinction is best captured via the path-integral representation of the density matrix, as discussed in the following App.~\ref{a.pathintegral}. Here we give an alternative picture solely based on the operator picture of the density matrix. In the following we shall choose, as observable of interest, the particle number $\hat{N}_A$ in the region $A$ of the system, capturing the quantum correlations between the region in question and its complement.

 When leaving the ground state of local Hamiltonians, one expects to encounter states with the generic feature of possessing volume-law entanglement, and volume-law fluctuations of particle number \cite{Page1993}. Hence one may naively suspect that, when dealing with excited states, the quantum coherent fluctuations are stronger, and not weaker, than in the ground state. This is indeed true, but it is only meaningful provided that, in an experiment, one is able to deterministically prepare one and the same excited state, in order to probe its fluctuation properties over many shots of the experiment itself. This last requirement is generally prohibitive, as experiments on excited states generally probe the properties of \emph{ensembles} (every shot of the experiment reproducing a different state). Whence the relevance of the concept of density matrices $\hat{\rho}$, not only in the context of systems coupled to dissipative baths, but also in the context of systems evolving uniquely under their own Hamiltonian dynamics. 
 
  In the latter case, let $|\Psi(t)\rangle$ be the instantaneous state of the system, and let $\Theta$ be a time window sufficiently long for time averages to equal ensemble averages (namely averages over repeated shots of the experiment). Then the density matrix describing the ensemble is well described by 
  \begin{equation}
  \hat{\rho} \approx \frac{1}{\Theta} \int_0^{\Theta} dt ~|\Psi(t)\rangle \langle \Psi(t)|
  \end{equation}
Despite the fact that each $|\Psi(t)\rangle$ state may exhibit volume-law entanglement and coherent fluctuations, the ensemble properties are quite different. Indeed, we can write $|\Psi(t)\rangle$ as 
\begin{equation}
|\Psi(t)\rangle = \sum_{N_A} \sum_{\{n_i \}_{N_A}} c_{\{ n_i \};N_A}(t)~ |\{ n_i \},N_A\rangle
\end{equation}
where $|\{ n_i \};N_A\rangle$ is a Fock state $\{ n_i \}$ characterized by having $N_A$ particle in $A$. After time/ensemble averaging, the density matrix takes the form
$\hat{\rho} = \hat{\rho}_{\rm D} + \hat{\rho}_{\rm OD}$ where
\begin{eqnarray}
\hat{\rho}_D = ~~~~~~~~~~~~~~~~ &  \\
\sum_{N_A} \sum_{\{n_i \}_{N_A}} \sum_{\{n_i' \}_{N_A}} &  \rho_{\{ n_i \},N_A;\{ n_i' \},N_A}  |\{ n_i \},N_A\rangle \langle \{ n'_i \},N_A |   \nonumber 
\end{eqnarray} 
is the diagonal part of the density matrix (in terms of the quantum number $N_A$), and 
\begin{eqnarray}
\hat{\rho}_{\rm OD} = ~~~~~~~~~~~~~~~~ &  \\
\sum_{N_A \neq N_A'} \sum_{\{n_i \}_{N_A}} \sum_{\{n_i' \}_{N_A'}} &  \rho_{\{ n_i \},N_A;\{ n_i' \},N_A'}  |\{ n_i \},N_A\rangle \langle \{ n'_i \},N_A' |   \nonumber 
\end{eqnarray} 
is the off-diagonal part; here
\begin{equation}
\rho_{\{ n_i \},N_A;\{ n_i' \},N_A'} = \frac{1}{\Theta} \int_0^{\Theta} dt ~ c_{\{ n_i \},N_A}(t) c^*_{\{ n_i' \},N'_A}(t)~.
\label{e.coherence}
\end{equation}
It is evident that $[\hat{\rho}_{\rm D},\hat{N}_A] = 0$ while $[\hat{\rho}_{\rm OD},\hat{N}_A] \neq 0$. Therefore the off-diagonal part, containing the coherence between configurations differing by the number of particles in $A$, is the part of $\hat\rho$ responsible for the quantum fluctuations of $\hat{N}_A$, captured by the quantum variance. { The total, extensive fluctuations of $N_A$ are given by the diagonal part, $\langle \delta^2 N_A \rangle = \langle \hat{N}_A^2 \rangle_{\hat{\rho}_D} - \langle \hat{N}_A \rangle^2_{\hat{\rho}_D}$; as a consequence the quantum coherent contribution, which stems from the off-diagonal terms, remains hidden in this calculation, and it cannot be formally separated from the incoherent part.} The proper separation between incoherent and coherent fluctuations is achieved within the path-integral formalism, as described in the main text and below in App.~\ref{a.pathintegral}. Nonetheless the operator form of the density matrix provides further insight into the physical origin and spatial structure of coherent quantum fluctuations, as discussed below.

The instantaneous coherence $c_{\{ n_i \},N_A}(t) c^*_{\{ n_i' \},N'_A}(t)$ connects states with $N_A-N_A' \sim {\cal O}(L_A^{d/2})$, as it is typical of excited states in Hilbert space. But the time/ensemble-averaged coherence $\rho_{\{ n_i \},N_A;\{ n_i' \},N_A'}$ in Eq.~\eqref{e.coherence} has a much shorter range away from the diagonal. Indeed, assuming that 
$\{ n_i \}$ and $\{ n_i' \}$ are connected by moving $m$ particles from sites $j_1, ..., j_m$ to sites $i_1, ..., i_m$, one has 
\begin{eqnarray} 
\rho_{\{ n_i \},N_A;\{ n_i' \},N_A'} = ~~~~~~~~~~~~~~&  \\
{\rm Tr}  \Big[ \hat{\rho}~ b^{\dagger}_{i_1} \cdot \cdot \cdot b^{\dagger}_{i_m} b_{j_1} \cdot \cdot \cdot b_{j_m}& |\{ n_i \},N_A\rangle \langle \{ n_i \},N_A | \Big ]
\nonumber
\end{eqnarray}
where $b_i, b_i^{\dagger}$ are the destruction/creation operators of the particles of interest (of arbitrary statistics). Hence, as one may have expected, the magnitude of $\rho_{\{ n_i \},N_A;\{ n_i' \},N_A'}$ is controlled by that of the $2m$-point correlation function, namely 
\begin{equation}
|\rho_{\{ n_i \},N_A;\{ n_i' \},N_A'}| \leq |\langle  b^{\dagger}_{j_1} \cdot \cdot \cdot b^{\dagger}_{j_m} b_{j_1} \cdot \cdot \cdot b_{j_m} \rangle| ~.
\end{equation}
In general such a correlation function will exhibit a fast decay with the (minimum) distances between pairs of points $i_p$ and $j_q$. This in turn implies that, in order to have a sizable coherence (Eq.~\eqref{e.coherence}), two  configurations $\{ n_i \}$ and $\{ n_i' \}$ should differ by particle moves which, when occurring between $A$ and its complement $B$, are localized (algebraically or exponentially) around the boundary between the two regions. This observation generally excludes a volume law for the coherent part of particle-number fluctuations, and it leaves an area law (up to multiplicative logarithmic corrections) as the only possibility.   

\section{Path-integral representation of a generic density matrix and of the quantum variance}
\label{a.pathintegral}

In this section we derive the path-integral representation for a generic density matrix, generalizing the discussion of the main text to arbitrary mixed states beyond thermal equilibrium. Moreover we derive the path-integral expression for the quantum variance. 

Any (semi-positive definite) density matrix $\hat{\rho}$ can always be cast in the form 
\begin{equation}
\hat\rho = \frac{e^{-\beta\hat{\cal H}}}{{\rm Tr}[e^{-\beta\hat{\cal H}}]},
\label{e.thermalrho}
\end{equation}
namely in the form of a thermal density matrix with (effective) temperature $k_B T = 1/\beta$. For generic (non-thermal) mixed states the specific value of $\beta$ is completely irrelevant, and one could set in the following $\beta = 1$ in some convenient energy units; yet, in order to make contact with the case of thermal equilibrium, hereafter we will keep the inverse temperature $\beta$ explicitly indicated.   
We consider a generic observable $\hat{O}$ which is diagonalized by a basis $|O_{\alpha},\{\bm \lambda\}_{\alpha}\rangle$, where $O_{\alpha}$ is the eigenvalue for $\hat{O}$, and $\{\bm \lambda\}_{\alpha}$ are the other quantum numbers possibly labeling the state. The partition function ${\cal Z} = {\rm Tr}[\exp(-\beta\hat{\cal H})]$ can be cast in the form of the trace of the product of infinitesimal propagators between successive states  $| O_{\alpha_i},\{\bm \lambda\}_{\alpha_i} \rangle$, namely
\begin{equation}
{\cal Z} = \lim_{M\to\infty} \sum_{\{\alpha_i \}} \prod_{i=1}^{M-1} \langle O_{\alpha_i},\{\bm \lambda\}_{\alpha_i} | e^{-\frac{\beta}{M}  \hat{\cal H} } | O_{\alpha_{i+1}},\{\bm \lambda\}_{\alpha_{i+1}}\rangle 
\end{equation} 
where $\sum_{\{\alpha_i \}}$ is a short-hand notation for the multiple sum over the quantum numbers  $ ( O_{\alpha_i},\{\bm \lambda\}_{\alpha_i})$ labeling each state in the propagation sequence $\alpha_1, \alpha_2, ..., \alpha_M\equiv\alpha_1$.
Summing over the ${\bm \lambda}$ quantum numbers, and taking the limit $M\to\infty$, one obtains formally the path-integral expression
\begin{equation}
{\cal Z} = \int_{O(0)\equiv O(\beta)} \hspace{-1cm} {\cal D}[O(\tau)] ~~e^{-S[O(\tau),\partial_{\tau}O(\tau),...]}
\end{equation} 
where $O(\tau)$ is the continuum limit of the sequence $\left \{ O_{\alpha_1},O_{\alpha_2},...,O_{\alpha_M} \right \}$, and 
\begin{equation}
e^{-S}  = \lim_{M\to\infty} \sum_{\{ {\bm \lambda}_{\alpha_i} \}} \prod_{i=1}^{M-1} \langle O_{\alpha_i},\{\bm \lambda\}_{\alpha_i} | e^{-\frac{\beta}{M}  \hat{\cal H} } | O_{\alpha_{i+1}},\{\bm \lambda\}_{\alpha_{i+1}}\rangle ~.
\end{equation}
Once the density matrix has been given the thermal form Eq.~\eqref{e.thermalrho}, it is straightforward to deform the density matrix upon application of a field $h$ coupling to $\hat O$, 
\begin{equation}
\hat\rho(h) = \frac{e^{-\beta(\hat{\cal H}-h\hat{O})}}{{\rm Tr}[e^{-\beta(\hat{\cal H}-h\hat{O})}]},
\label{e.thermalrho_def}
\end{equation}
which allows one to define the response function in the standard way as $\chi_O = \frac{\partial}{\partial h} {\rm Tr}[ \hat\rho(h) \hat{O}] \Big|_{h=0}$.

The path-integral representation of response function leads to the expression
\begin{eqnarray}
\chi_O & = & \left \langle \int d\tau  ~\delta O(\tau) \delta O(0) \right \rangle_S   \nonumber \\
 & = & \frac{1}{\beta} \left \langle \int d\tau \int d\tau'  ~\delta O(\tau) \delta O(\tau') \right \rangle_S
\label{e.PIthermfluct}
\end{eqnarray}
where $\delta O(\tau) = O(\tau) - \langle O \rangle_S$, and we have invoked the periodicity of $O(\tau)$ paths in imaginary time. Here 
\begin{equation}
\langle ... \rangle_S =  \frac{1}{{\cal Z}} \int_{O(0)\equiv O(\beta)}\hspace{-1cm} {\cal D}[O(\tau)] ~  (...)~ e^{-S}
\end{equation}
is the average over the space of $O(\tau)$ paths. Moreover one has that 
\begin{equation}
\langle \delta^2 O \rangle  =  \frac{1}{\beta} \left \langle \int d\tau  ~(\delta O(\tau))^2  \right \rangle_S ~.
\label{e.PItotfluct}
\end{equation}
Combing Eqs.~\eqref{e.PIthermfluct} and \eqref{e.PItotfluct}, one readily obtains the path-integral expression for the quantum variance 
\begin{eqnarray}
\langle \delta^2 O \rangle_Q & = & \langle \delta^2 O \rangle - \chi_O~ k_B T \nonumber  \\
&= & \frac{1}{\beta} \left \langle \int d\tau \left[ O(\tau) - \frac{1}{\beta} \int d\tau' ~O(\tau') \right]^2 \right \rangle_S
\end{eqnarray}
showing that it represents the average variance of fluctuations of $O(\tau)$ paths around their centroid. 

 {We end this section by noticing that the deformation of the density matrix to Eq.~\ref{e.thermalrho_def} is a physically meaningful operation for thermal states - as it can be obtained by turning on the perturbation $-h\hat{O}$ in the Hamiltonian within an isothermal setting -- see App.~\ref{a.exp} for further discussion in the specific case of quantum gas microscopes. Hence in the case of thermal states, neither the measurement nor the calculation of the quantum variance requires the full knowledge of the density matrix. On the other hand, for generic mixed states the deformation of $\hat{\rho}$ should be thought of in general as a mathematical operation. Devising physical  (namely, experimentally realistic) operations that can lead to the deformation of a generic density matrix as in Eq.~\ref{e.thermalrho_def} is an outstanding task, which we postpone to future investigations.}

\begin{center}
  \begin{figure}
  \includegraphics[width=8cm]{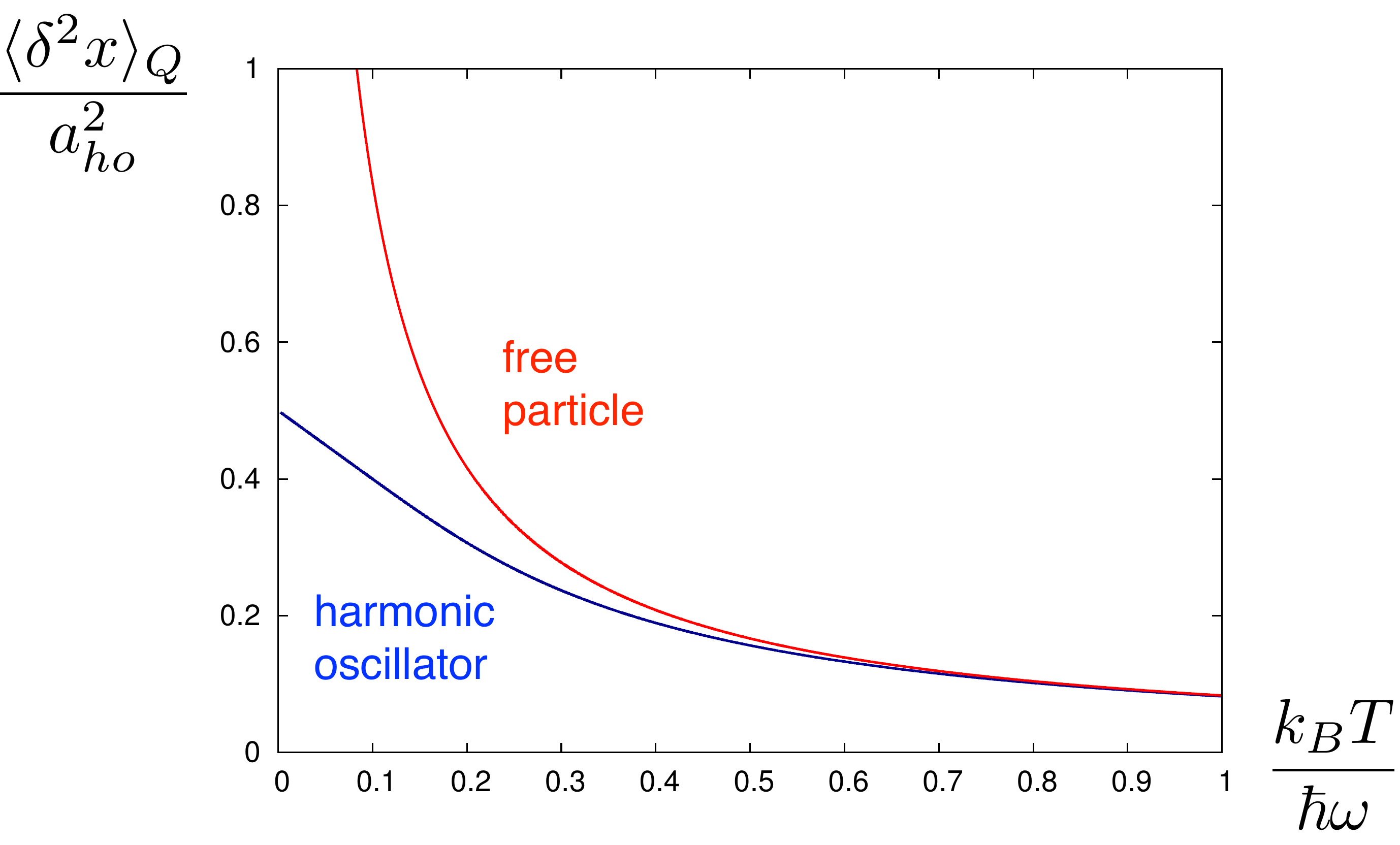}
  \caption{{\bf Quantum variance of the position of a 1$d$ particle.} The plot shows the quantum variance of the position for a one-dimensional harmonic oscillator, as well as for a free particle as a function of temperature. In the case of the free particle the frequency $\omega$ is to be understood as an arbitrary constant setting the energy scale $\hbar \omega$ and length scale $a_{\rm ho}$.}
\label{f.qvsimple}
\end{figure}
\end{center}

\section{Quantum variance as generalized de Broglie wavelength}
\label{a.deBroglie}

\subsection{Quantum variance for simple models}
The analytical calculation of the quantum variance of the position operator is illuminating in the case of simple models, namely the free particle and the harmonic oscillator in one dimension. It is convenient to start from the second one, and to obtain the free-particle result as a limiting case. 
In the case of the harmonic oscillator, the position fluctuations are readily obtained from the diagonal part of the density matrix $\langle x | e^{-\beta \hat{\cal H}} |x \rangle/{\cal Z}$ \cite{FeynmanHibbs}, while the susceptibility $\chi_x = \frac{\partial \langle x \rangle}{\partial h}$ to a displacement  of the harmonic oscillator potential $\frac{1}{2} m\omega x^2 \to \frac{1}{2} m\omega x^2 - h x$ is readily obtained by the linear displacement of the average, $\langle x \rangle \to \langle x \rangle - h/(m\omega^2)$. As a result the quantum variance takes the form
\begin{equation}
\langle \delta^2 x \rangle_Q = \frac{a^2_{\rm ho}}{2} \left[ \frac{\sinh(1/\theta)}{\cosh(1/\theta)-1} - 2 \theta \right ] ~~~~~ {\rm (harm. ~osc.)}
\label{e.qvho}
\end{equation}
where $a_{\rm ho} = \sqrt{\hbar/(m\omega)}$ and $\theta = k_B T / \hbar \omega$. In the limit $T\to 0$ one recovers Heisenberg's uncertainty in the ground state, $\langle \delta^2 x \rangle_0 = a_{ho}^2/2$. 

On the other hand, taking the limit $\omega\to 0$ gives the result for the free particle, which, after careful expansion of Eq.~\eqref{e.qvho}, gives
\begin{equation}
\langle \delta^2 x \rangle_Q = \frac{1}{24 \pi}~ \lambda^2_{dB}(T) ~~~~~~~ {\rm (free~ particle)}
\end{equation} 
where $\lambda_{dB}(T) = \sqrt{2\pi \hbar^2/ (m k_B T)}$ is the thermal de Broglie wavelength. 
The link between the quantum variance and the de Broglie wavelength shows that the quantum variance of the particle position gives the characteristic (squared) amplitude of coherent quantum fluctuations at finite temperature \cite{FeynmanHibbs}.  In fact one may interpret the quantum variance of the position as a generalization of the thermal de Broglie wavelength to the case of a particle in an external potential, such as the case of the harmonic oscillator. In Fig.~\ref{f.qvsimple} the quantum variance of the position for the two models discussed above shows the expected monotonic decrease with temperature, due to the shrinking of the imaginary-time ``duration" of coherent quantum fluctuations.
 
\begin{center}
  \begin{figure}
  \includegraphics[width=\columnwidth]{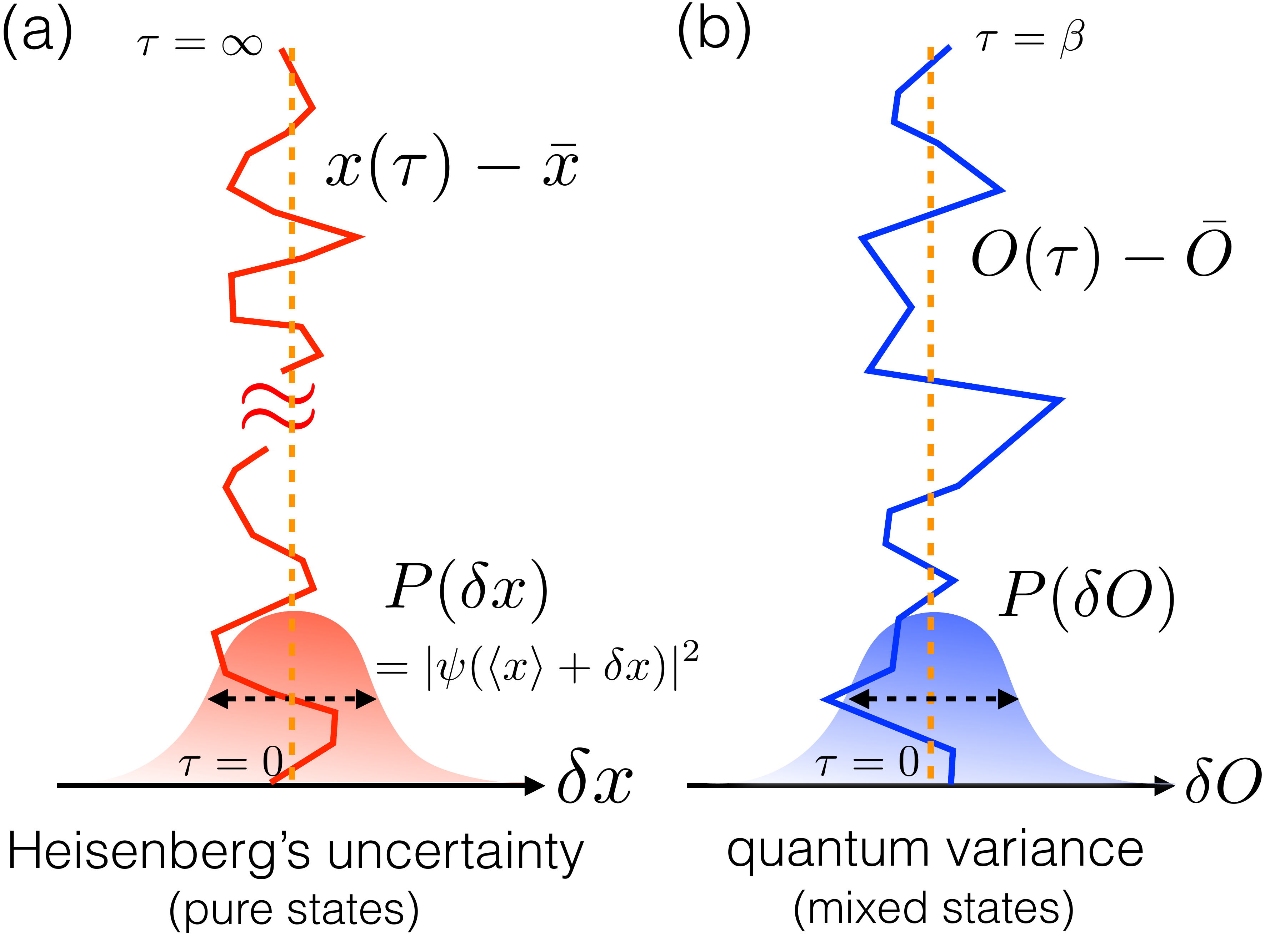}
  \caption{{\bf Quantum variance as Heisenberg's uncertainty for mixed states.} (a) The path-integral representation of the partition function for a single one-dimensional particle at $T\to 0$ (or for a generic pure state) is a sum over infinitely long paths in imaginary time $x(\tau)$; on average each path visits the region $[x,x+dx]$ a number of times proportional to $|\psi(x)|^2 dx$, where $\psi(x)$ is the ground-state (or, more generally, the pure-state) wavefunction. The width of the wave function's square modulus gives the Heisenberg's uncertainty, namely the amplitude of coherent quantum fluctuations; (b) In the case of mixed states of a generic many-body system, the probability distribution for the coherent quantum fluctuations of a generic observable $\hat{O}$ is instead given by the probability that the (imaginary-time) instantaneous value of $O(\tau)$ differs from the path centroid $\bar{O} = \beta^{-1} \int d\tau O(\tau)$.}
  \label{f.QV_Heisenberg}
  \end{figure}
  \end{center}

\subsection{Imaginary-time fluctuations as quantum coherent fluctuations}

As seen in App.~\ref{a.pathintegral}, the quantum variance of a generic observable gives the characteristic amplitude of fluctuations for such an observable along the imaginary-time dynamics of the system. On the other hand, in App.~\ref{a.deBroglie} we have established a direct relationship between the quantum variance of a free particle and the thermal de Broglie wavelength, namely the characteristic width of wavepackets at finite temperature. In this section we bring the two observations together to argue that the quantum variance generalizes the concept of thermal de Broglie wavelength, or finite-temperature coherence length, to the space of eigenvalues of \emph{any} observable (not only the position operator) and for \emph{any} quantum system.  

The relationship between the quantum variance and the de Broglie wavelength is very natural when considering the fundamental link existing between the wavefunction of a pure state and the statistics of Feynman paths.  
Thanks to the parametrization in Eq.~\eqref{e.thermalrho}, a pure state can always be thought of as the $\beta\to\infty$ limit of a mixed-state density matrix, and hence represented in the form of a path integral. In the case of a one-dimensional quantum particle, the path integral for a pure state with wavefunction $\psi(x)$ runs over infinitely long trajectories $x(\tau)$, whose fluctuations $\delta x$ around the average position $\langle x \rangle = \langle \psi | {\hat x} |\psi \rangle$ obey the statistics dictated by the modulus of the square function \cite{RossiT1983}, namely, taking an arbitrary time $\tau$: 
\begin{eqnarray}
P(\delta x) &= &  \frac{1}{\cal Z} \int {\cal D}[x(\tau)] ~ \delta (x(\tau)-\langle \hat x \rangle-\delta x) ~e^{-S} \nonumber \\
& = & |\psi(\langle x \rangle+\delta x)|^2 ~.
\end{eqnarray}
{In particular any infinite trajectory contributing to the path integral has the same statistical properties as the whole ensemble, so that the centroid of the path $\bar x = \bar x[x(\tau)]$ must correspond to the expectation value $\langle \hat x \rangle$.}  Hence, as depicted in Fig.~\ref{f.QV_Heisenberg}(a), the imaginary-time fluctuations span the support of the wavefunction, and the (squared) amplitude of fluctuations of the path $x(\tau)$ around its centroid $\bar x$ -- the quantum variance -- is the same as the (squared) width of the wavefunction, giving Heisenberg's uncertainty.  
In the case of free particles, the thermal de Broglie wavelength generalizes Heisenberg's uncertainty on the position to the case of thermal states. Therefore, it is not too surprising that the quantum variance follows the de Broglie wavelength at finite temperature, as shown in App.~\ref{a.deBroglie}.

 The concept of quantum variance extends all the above considerations to generic observables and generic quantum systems. The quantum variance provides the width of the probability distribution for the fluctuations of generic observables around the path centroid (see Fig.~\ref{f.QV_Heisenberg}(b)), namely
 \begin{equation}
 P(\delta O) =  \frac{1}{\cal Z} \int {\cal D}[O(\tau)] ~ \delta (O(\tau) - \bar{O} - \delta O) ~e^{-S}
 \end{equation}
 where $\bar{O} = \bar{O}[O(\tau)] $ is the centroid. As seen in the case of the position of a one-dimensional particle, for a pure state the width of $P(\delta O)$ expresses the Heisenberg's uncertainty on the observable $\hat{O}$. When applied to a mixed state, the quantum variance generalizes therefore Heisenberg's uncertainty, expressing the (squared) amplitude of coherent quantum fluctuations of the observable.

\section{Quantum variance vs. skew information and local quantum uncertainty}
\label{a.skewinfo}

In this section we shall discuss the relationship between the quantum variance and the skew information \cite{WignerY1963}, the latter being a widespread concept in quantum information to quantify the quantum uncertainty of an observable. In particular we shall show that the quantum variance provides a tight lower bound, based on physical observables, to the otherwise abstract skew information. Moreover the discussion of the relationship between the skew information and the quantum variance allows one to conclude on the convexity of the latter. Finally we will see how the quantum variance relates to the recently introduced ``local quantum uncertainty" \cite{Girolamietal2013}, which is advocated as a measure of quantum correlations.     

\subsection{Wigner, Dyson, Lieb and the convexity of quantum variance}

The Dyson-Wigner-Yanase (DWY) \emph{skew information} \cite{WignerY1963,Wehrl1978} 
\begin{equation}
I_{\alpha}(\hat{O},\hat{\rho}) = - \frac{1}{2} {\rm Tr} \{ [\hat{O},\hat\rho^{\alpha}][\hat{O},\hat{\rho}^{1-\alpha}] \}~~~~ (0\leq \alpha \leq 1)
\end{equation}
probes the quantum uncertainty of the observable $\hat{O}$ due to its non-commutativity with the density matrix of the system. Replacing $\hat O$ by $\delta \hat O = \hat O - \langle \hat O \rangle$ does not alter the above definition. 

Writing again the generic density matrix $\hat{\rho}$ as a thermal state, $\hat\rho = e^{-\beta \hat{\cal H}}/{\cal Z}$ (with an arbitrary effective inverse temperature $\beta$), one can immediately show that the DWY skew information can be expressed as an imaginary-time correlation function 
\begin{equation}
I_{\alpha}(O,\rho) = \langle \delta^2 \hat{O} \rangle - \langle \delta \hat{O}(\tau=\alpha\beta) \delta \hat{O}(0) \rangle ~.
\label{e.skew_beta2}
\end{equation}
Hence, clearly, the DWY skew information $I_{\alpha}$ expresses the amount by which  the imaginary-time correlation function $\langle \delta \hat{O}(\tau) \delta \hat{O}(0) \rangle$ at a time $\tau = \alpha \beta$ has decreased with respect to the equal-time ($\tau=0$) value. Hence the DWY skew information probes the imaginary time fluctuations in a similar manner to quantum variance. As a consequence a link between the two quantities can be expected, and it is straightforwardly established in the form 
\begin{eqnarray}
\langle \delta^2 \hat{O} \rangle_Q & = & \langle \delta^2\hat{O} \rangle - \frac{1}{\beta} \int_0^{\beta} d\tau ~\langle \delta\hat{O}(\tau) \delta\hat{O}(0) \rangle \nonumber \\
 & =  & \int_0^1 d\alpha ~ I_{\alpha}(\hat{O},\hat{\rho})
\label{e.qv_si}
\end{eqnarray}
namely the quantum variance is equal to the \emph{average DWY skew information}. In particular the Wigner-Yanase (WY) skew information \cite{WignerY1963},
$I_{\alpha=1/2}$, is an upper bound to the DWY skew information
\begin{equation}
I_{\alpha}(\hat{O},\hat{\rho}) \leq I_{\alpha=1/2}(\hat{O},\hat{\rho}) 
\end{equation}
as it is easy to prove due to the monotonic decay of imaginary-time correlation functions, $\langle \delta\hat{O}(\tau) \delta\hat{O}(0) \rangle \geq \langle \delta\hat{O}(\tau=\beta/2) \delta\hat{O}(0) \rangle$. As a consequence one readily obtains that the quantum variance is always lower than the WY skew information
\begin{equation}
\langle \delta^2 \hat{O} \rangle_Q \leq I_{\alpha=1/2}(\hat{O},\hat{\rho})~.
\label{e.qv_WY}
\end{equation}

Lieb \cite{Lieb1973} proved that the DWY skew information is \emph{convex} for any value of $\alpha$, namely:
\begin{equation}
I_{\alpha}(\hat{O},\lambda_1 \hat{\rho}_1 + \lambda_2 \hat{\rho}_2) \leq \lambda_1  I_{\alpha}(\hat{O},\hat{\rho}_1) +  \lambda_2  I_{\alpha}(\hat{O},\hat{\rho}_2)
\end{equation}
for any real numbers $\lambda_1$, $\lambda_2$.
Using Eq.~\eqref{e.qv_si}, the property of convexity is immediately inherited by the quantum variance. { The convexity of quantum variance is a fundamental figure of merit to assess the quantum variance as a probe of quantum coherent fluctuations: if $\hat\rho_1$ and $\hat\rho_2$ have the same quantum variance, any linear incoherent superposition of the two has necessarily a lower quantum variance.} 

\subsection{Quantum variance vs. local quantum uncertainty}
\label{s.QV_LQU}

Given an $A$-$B$ bipartition of an extended quantum system, Ref.~\cite{Girolamietal2013} has introduced the concept of \emph{local quantum uncertainty} (LQU) 
 \begin{equation}
 {\cal U}_A^{\Lambda}(\hat\rho) = {\rm min}_{\hat{O}^{\Lambda}_A}~ I_{1/2}(\hat{O}^{\Lambda}_A,\hat\rho)
 \label{e.LQU}
 \end{equation} 
as the minimum of the WY skew information over all local observables $O^{\Lambda}_A$ in $A$ having a given spectrum $\Lambda$. Ref.~\cite{Girolamietal2013} argues that this observable-independent (but spectrum-dependent) quantity acts as a discord-like quantity, namely a measure of quantum correlations between $A$ and $B$. In order to capture quantum correlations among \emph{all}, equally weighted degrees of freedom of $A$ and those of $B$, it is obvious to request that the observable $O_A$ be an extensive one, namely the sum of local observables $\hat{O}_j$, 
\begin{equation}
\hat{O}_A^{\rm (macro)} = \sum_{j\in A} \hat{O}_j~.
\label{e.sum}
\end{equation}
 This ensures that quantum fluctuations of all degrees of freedom in $A$ are taken into account into the skew information. As the size of $A$ grows, the spectrum of the operator $\hat{o}_A^{\rm (micro)} = \hat{O}^{\rm (macro)}_A/L_A^d$ becomes a continuous one, and it is contained in a finite interval $[\lambda_{\rm min}, \lambda_{\rm max}]$. This behavior applies to all extensive operators of the kind of $\hat{O}_A^{\rm (macro)}$, and their spectrum can easily be reduced to one and the same $\Lambda$ by a simple shift and rescaling in the definition of the operator. Hence, in the sense of Ref.~\cite{Girolamietal2013}, one can define a \emph{macroscopic} LQU ${\cal U}_A^{\rm (macro)}(\rho)$  defined as a minimum over all operators $\hat{O}_A^{\rm (macro)}$, which is arguably the most appropriate definition of discord-type correlations among all degrees of freedom of $A$ and those in $B$. {We assume that $A$ and $B$ interact with a coherent Hamiltonian term, leading to an exchange of energy, and possibly also particle, or magnetization, etc... Hence in the minimization procedure we explicitly exclude the possible existence of local conserved quantities $[\hat{O}^{\rm (macro)}_A, \hat{\rho}]=0$, which would trivially lead to a vanishing macroscopic LQU. }
 
  It follows immediately from Eq.~\eqref{e.qv_WY} that the macroscopic LQU is lower-bounded by the minimum quantum variance of macroscopic observables
 \begin{eqnarray}
 {\cal U}_A^{\rm (macro)}(\hat\rho) & = & {\rm min}_{\hat{O}^{\rm (macro)}_A}~ I_{1/2}(\hat{O}^{\rm (macro)},\hat\rho) \nonumber \\
  & \geq &  {\rm min}_{\hat{O}^{\rm (macro)}_A}~ \langle \delta^2  \hat{O}^{\rm (macro)}_A \rangle_Q~.~~~~~
  \label{e.macroLQU}
  \end{eqnarray} 
 The minimization implied by Eq.~\eqref{e.macroLQU} is readily performed for the quantum variance: the minimum quantum variance of macroscopic observables is realized by bipartite fluctuations of an otherwise conserved quantity, namely $\hat{O}_A$ such that $[\hat{O}_A + \hat{O}_B,\rho]=0$. For general quantum systems, the above requirement applies to the local energy, and, in the presence of a continuous symmetry, to the local particle number (for particle models) or to the local magnetization (for spin models), etc. -- assuming that the latter quantities are not conserved. In the case of equilibrium states of local Hamiltonians, we have shown in this work that the quantum variance of bipartite fluctuations obeys an area law: as a consequence,  Eq.~\eqref{e.macroLQU} implies that the macroscopic LQU obeys \emph{at least} an area law. On the other hand, in the ground state the WY skew information reduces to the variance of the operator
 \begin{equation}
 I_{1/2}(\hat{O}_A^{\rm (macro)},\hat\rho) \underrel{T=0}{=}  \langle \delta^2 O_A^{\rm (macro)} \rangle~,
 \end{equation}
 and the scaling of the minimum variance of local macroscopic operators in the ground state of local Hamiltonians satisfies a (logarithmically violated) area law \cite{FrerotR2015}. { Even though the temperature dependence of the WY skew information is not generally known in the literature, one can assume that it is maximized at $T=0$ (this is the case of the free-fermion example studied explicitly in Sec~\ref{s.correlations}). Under this assumption, and given that the WY skew information at $T=0$ coincides with the total variance, we obtain the inequalities}
 \begin{eqnarray}
 &&   {\rm min}_{\hat{O}^{\rm (macro)}_A}~ \langle \delta^2  \hat{O}^{\rm (macro)}_A \rangle_Q  \\
 & \leq & {\cal U}_A^{\rm (macro)}(\hat\rho)  \leq 
 {\rm min}_{\hat{O}^{\rm (macro)}_A}~ \langle \delta^2 O^{\rm (macro)} \rangle (T=0) \nonumber
 \end{eqnarray}
 implying that the macroscopic LQU obeys \emph{at most} a logarithmically violated area law, namely
  \begin{equation}
 {\cal O}(L_A^{d-1}) \leq {\cal U}_A^{\rm (macro)}(\hat\rho) \leq {\cal O}(L_A^{d-1}\log L_A )~.
 \label{e.LQUscaling}
 \end{equation}
 
\section{Quantum variance vs. quantum Fisher information: quantum correlations and metrology}
\label{a.QFI}

In this section we focus on the relationship between the quantum variance and the quantum Fisher information \cite{BraunsteinC1994}, a central quantity in quantum metrology due to its link with the maximum precision achievable in the estimation of the parameter of a given unitary transformation. Similarly to the skew information, the quantum variance offers a lower bound to the quantum Fisher information; we shall exploit this fact in the context of the recently introduced ``interferometric power" \cite{Girolamietal2014} to explore the importance of the quantum variance of bipartite fluctuations both for metrology and for quantum correlations. Further implications of this bound in the context of entanglement witnessing will be discussed in Sec~\ref{a.enta}. 

\subsection{Quantum variance as a lower bound to the quantum Fisher information}

The quantum Fisher information (QFI) \cite{BraunsteinC1994} expresses the ``distinguishability" (in the sense of the Bures distance) between two density matrices $\hat{\rho}(h)$ and $\hat{\rho}(h+\delta h)$, belonging to a family $\hat{\rho}(h)$ continuously parametrized by the parameter $h$. If the family of density matrices is obtained via a unitary transformation generated by an Hermitian operator $\hat{O}$, $\hat{\rho}(h) = e^{-i\hat{O}h}\hat{\rho}(h=0) e^{i\hat{O}h}$, the QFI takes the explicit form $F_Q(\hat{O},\hat{\rho}) =  \sum_{nm}  G_{\rm F}(p_n,p_m)~|\langle n | \delta \hat{O} |m\rangle|^2$ where $p_n$ and $| n \rangle$ are eigenvalues and eigenvectors of the density matrix, and 
\begin{equation}
G_{\rm F}(p_n,p_m) = 2 \frac{(p_n - p_m)^2}{p_n+p_m}~.
\end{equation}
This is to be compared with the expression of the quantum variance, namely $\langle \delta^2 O \rangle_Q = \sum_{nm}  G_{\rm QV}(p_n,p_m) |\langle n | \delta O | m \rangle|^2$ where
\begin{equation}
  G_{\rm QV}(p_n,p_m) = \frac{p_n+p_m}{2} - \frac{p_n-p_m}{\log(p_n)-\log(p_m)}~.
  \end{equation}
  Comparing the two functions it is easy to realize that 
  \begin{equation}
  \frac{G_{\rm F}(x,y)}{4} \geq   G_{\rm QV}(x,y) ~~~~~~ 0 \leq x,y \leq 1
  \end{equation}
  whence the announced inequality
  \begin{equation}
  \frac{F_Q(\hat{O},\hat{\rho})}{4} \geq \langle \delta^2 \hat O \rangle_Q~.
  \label{e.qv_qfi}
  \end{equation}
  {
  Incidentally we notice that the WY skew information admits a similar expression $I_{1/2}(\hat{O},\hat{\rho}) =  \sum_{nm}  G_{{\rm I}_{1/2}}(p_n,p_m)~|\langle n | \delta \hat{O} |m\rangle|^2$ with
  \begin{equation}
G_{{\rm I}_{1/2}}(p_n,p_m) = \frac{p_n+p_m}{2} - \sqrt{p_n p_m}~.
\end{equation}
Direct inspection into the $G$ functions reveals the inequality chain:
 \begin{equation}
 \frac{G_{\rm F}(x,y)}{4} \geq   G_{{\rm I}_{1/2}}(x,y) \geq G_{\rm QV}(x,y)  ~~~~~~ 0 \leq x,y \leq 1
\end{equation}
whence the ensuing hierarchy:
 \begin{equation}
  \frac{F_Q(\hat{O},\hat{\rho})}{4} \geq I_{1/2}(\hat{O},\hat{\rho}) \geq \langle \delta^2 \hat O \rangle_Q~.
  \label{e.hierarchy}
  \end{equation}
 In particular we notice that QFI and WY skew information have often been invoked as sharing similar properties \cite{Luo2004}, yet a well-defined relationship as the inequality of Eq.~\eqref{e.hierarchy} between the two has not yet been established (in this respect, Ref.~\cite{Luo2004} uses a QFI-like quantity, which is not equivalent to the actual definition of QFI).  The quantum variance further explicits this relationship by finding a common, non trivial lower bound for both quantities. }

\subsection{Metrological implications: the interferometric power}

In a similar manner to the definition of local quantum uncertainty discussed in App.~\ref{s.QV_LQU}, Ref.~\cite{Girolamietal2014} has introduced the concept of \emph{interferometric power} (IP) of an observable $O_A^{\Lambda}$ in a bipartite ($A+B$) system with density matrix $\hat{\rho}$ as 
\begin{equation}
{\cal P}_A^{\Lambda}(\hat\rho) = \frac{1}{4} \min_{\hat{O}_A^{\Lambda}} F_Q(\hat{O}_A^{\Lambda},\hat{\rho})
 \end{equation}
 where $F_Q(\hat{O}_A^{\Lambda},\hat{\rho})$ is the quantum Fisher information associated with a unitary transformation generated by a local observable $\hat O_A^{\Lambda}$ acting on $A$, and with spectrum $\Lambda$. The IP has a direct metrological meaning: it expresses the worst-case-scenario uncertainty (in the sense of the Cram\'er-Rao bound \cite{Helstrom1976}) that one can achieve in the estimation of the parameter of a unitary transformation generated by an arbitrary observable which is local in $A$ and has a given spectrum ${\Lambda}$.  Ref.~\cite{Girolamietal2014} argues that the IP is another discord-type measure of quantum correlations between $A$ and $B$, leading to the conclusion that quantum correlations are a resource for metrology. 
 
 It is immediate to see that the above conclusions carry automatically over to the case of the quantum variance. Using the inequality Eq.~\eqref{e.qv_qfi}, one immediately has that 
 \begin{equation}
{\cal P}_A^{\Lambda}(\rho) \geq \min_{\hat{O}_A^{\Lambda}} \langle \delta^2 \hat{O}_A^{\Lambda} \rangle_Q~.
\label{e.QV_IP}
 \end{equation}
In App.~\ref{s.QV_LQU} we argued that macroscopic observables $\hat{O}_A^{\rm (macro)}$ in $A$, having an extensive spectrum $\Lambda$, capture the quantum correlations between all degrees of freedom in $A$ and those in $B$. In the case of such observables one can perform the minimization immediately for the right-hand side of Eq.~\eqref{e.QV_IP}, identifying the $O_A^{\rm (macro)}$ operator with the one satisfying the condition $[\hat{O}_A^{\rm (macro)}+\hat{O}_B^{\rm (macro)},\hat\rho]=0$ {(again, as in App.~\ref{s.QV_LQU} we are excluding local conserved quantities from the minimization)}. Hence the quantum variance of bipartite fluctuations provides a lower bound on the IP of macroscopic observables, and on the quantum correlations and metrological resource that this quantity expresses. { Similarly to Eq.~\eqref{e.LQUscaling}, this lower bound allows one to establish an area law scaling (with at most logarithmic corrections) to the IP of macroscopic observables, under the assumption (verified \emph{e.g.} by free fermions as in Sec.~\ref{s.correlations}) that the QFI is maximised at $T=0$}. In particular this bound is very instructive in terms of the metrological utility of many-body states: the maximum quantum variance of bipartite fluctuations, and hence the maximum IP, is achieved for states exhibiting power-law correlations, and specifically in the vicinity of quantum critical points  -- see also \cite{Haukeetal2015} for a recent calculation of the quantum Fisher information in exactly solvable models of quantum-critical points, which confirms this conclusion.

\section{Quantum variance, skew information and quantum Fisher information of bipartite fluctuations for free fermions}
\label{a.bipartitefluctu}

\subsection{Quantum variance}
 In this section we calculate the quantum variance of local particle-number fluctuations in the case of free fermions on a $d$-dimensional hypercubic lattice at half filling. 
 The density-density correlation function is given by 
 \begin{eqnarray}
 \langle \delta \hat{n}_i(\tau) \delta \hat{n}_j (0) \rangle = ~~~~~~~~~~~~~~&  \\
 \frac{1}{L^{2d}} \sum_{{\bm k},{\bm k}'}  e^{i ( {\bm k}-{\bm k}'  ) \cdot ( {\bm r}_i-{\bm r}_j  ) }  & e^{(\epsilon_{\bm k} - \epsilon_{\bm k}')\tau} f_{\bm k}\left ( 1- f_{{\bm k}'} \right )  \nonumber  
 \end{eqnarray}  
 where $f_{\bm k} = [\exp(\beta\epsilon_{\bm k})+1]^{-1}$ is the Fermi distribution, and $\epsilon_{\bm k} = -2J \sum_{\alpha = x, y, ...} \cos(k_{\alpha})$ is the dispersion relation. 
Integrating the correlation function to get $\langle \delta^2 \hat{N}_A \rangle$ and $\langle \delta^2 \hat{N}_A \rangle_T$, one obtains the quantum variance in the form
 \begin{eqnarray}
 \langle \delta^2 \hat{N}_A \rangle_Q   = ~~~~~~~~~~~~~~~~~~~~&  \\
 \frac{1}{L^{2d}} \sum_{{\bm k},{\bm k}'}  \sum_{i,j\in A} e^{i ( {\bm k}-{\bm k}'  ) \cdot ( {\bm r}_i-{\bm r}_j  ) }  &  f_{\bm k}\left ( 1- f_{{\bm k}'} \right ) 
 \left[ 1+ \frac{1 - e^{\beta (\epsilon_{\bm k} - \epsilon_{{\bm k}'})}}{\beta (\epsilon_{\bm k} - \epsilon_{{\bm k}'})} \right ]   \nonumber
 \end{eqnarray}  
In the high-temperature limit $\beta\to 0$ the quantum variance reduces to 
 \begin{eqnarray}
 \langle \delta^2 \hat{N}_A \rangle_Q   = ~~~~~~~~~~~~~~~~~~~~~~~~~~~&  \\
  \frac{\beta^2}{48} \frac{1}{L^{2d}} \sum_{{\bm k},{\bm k}'}  \sum_{i,j\in A} e^{i ( {\bm k}-{\bm k}'  ) \cdot ( {\bm r}_i-{\bm r}_j  ) } &
 (\epsilon_{\bm k} - \epsilon_{\bm k'})^2 + {\cal O}(\beta^3)~. \nonumber
 \label{e.QVhighT}
\end{eqnarray}
One observes that the term linear in $\beta$ vanishes, so that the dominant temperature dependence goes like $T^{-2}$. 

Assuming for $A$ the geometry of a hypercube of side $L_A$, the double sum over the $A$ region can be performed exactly, leading to 
 \begin{eqnarray}
 \langle \delta^2 \hat{N}_A \rangle_Q   =  ~&  \\
 \frac{\beta^2}{48} \left (\frac{L_A}{L} \right )^{2d} \sum_{{\bm k},{\bm k}'} &  \left( \prod_{\alpha = x, y,...}
 \frac{ {\rm sinc}^2[(k_\alpha-k_\alpha')L_A/2]}{{\rm sinc}^2[(k_\alpha-k_\alpha')/2]}  \right) \nonumber
 (\epsilon_{\bm k} - \epsilon_{\bm k'})^2 \\ 
 +~ {\cal O}(\beta^3) ~.~~~~~~~~~~~&~ \nonumber
\end{eqnarray}

In the limit $L_A\to \infty$ one has that 
\begin{equation}
 L_A \frac{ {\rm sinc}^2[(k_\alpha-k_\alpha')L_A/2]}{{\rm sinc}^2[(k_\alpha-k_\alpha')/2]}  \approx 2  \sum_{n=-\infty}^{\infty} \delta \left(k_\alpha-k'_\alpha - 2\pi n \right)~.
\end{equation}
One sees immediately that this limit would highlight a term scaling as $L_A^d$ (volume law), but that this term is actually vanishing because 
$\epsilon_{\bm k} = \epsilon_{\bm k + 2\pi n {\bm e}_{\alpha}}$ . Hence one is left with an area-law scaling term. 

\subsection{Wigner-Yanase skew information}

Using Eq.~\eqref{e.skew_beta2} leads immediately to the following expression for the skew information of bipartite particle-number fluctuations of free fermions
 \begin{eqnarray}
I_{1/2}(\hat{N}_A,\hat{\rho})  = ~~~~~~~~~~~~~~~~~~~~&  \\
 \frac{1}{L^{2d}} \sum_{{\bm k},{\bm k}'}  \sum_{i,j\in A} e^{i ( {\bm k}-{\bm k}'  ) \cdot ( {\bm r}_i-{\bm r}_j  ) }  &  f_{\bm k}\left ( 1- f_{{\bm k}'} \right ) 
 \left[ 1 -  e^{\beta (\epsilon_{\bm k} - \epsilon_{{\bm k}'})/2}\right ]   \nonumber
 \end{eqnarray}  
 which, when expanded at high temperature, leads to the behavior
\begin{eqnarray}
I_{1/2}(\hat{N}_A,\hat{\rho}) =  ~~~~~~~~~~~~~~~~~~~~~~~~~&  \\
  \frac{\beta^2}{32} \frac{1}{L^{2d}} \sum_{{\bm k},{\bm k}'}  \sum_{i,j\in A} e^{i ( {\bm k}-{\bm k}'  ) \cdot ( {\bm r}_i-{\bm r}_j  ) } & 
 (\epsilon_{\bm k} - \epsilon_{\bm k'})^2 + {\cal O}(\beta^3)~.  \nonumber
\end{eqnarray}
The above expression is very similar to Eq.~\eqref{e.QVhighT} for the quantum variance, confirming that the two quantities have the same high-temperature behavior (as well as the same zero-temperature value). In particular, when $\beta\to 0$:
 \begin{equation}
 \langle \delta^2 \hat{N}_A \rangle_Q  =   \frac{2}{3}~ I_{1/2}(\hat{N}_A,\hat{\rho}) +  {\cal O}(\beta^3)~.
 \end{equation}

\subsection{Quantum Fisher information}

Finally, when considering the QFI for free fermions \cite{Haukeetal2015}, one finds the following expression for bipartite particle-number fluctuations
 \begin{eqnarray}
F_Q(\hat{N}_A,\hat{\rho})    = ~~~~~~~~~~~~~~~~~~~~~~~&  \\
 \frac{4}{L^{2d}} \sum_{{\bm k},{\bm k}'}  \sum_{i,j\in A} e^{i ( {\bm k}-{\bm k}'  ) \cdot ( {\bm r}_i-{\bm r}_j  ) }  &  f_{\bm k}\left ( 1- f_{{\bm k}'} \right ) 
\tanh^2 \left[ \frac{\beta (\epsilon_{\bm k} - \epsilon_{{\bm k}'})}{2} \right ]    \nonumber
 \end{eqnarray}  
which leads to the high-temperature behavior 
\begin{eqnarray}
 F_Q(\hat{N}_A,\hat{\rho})   = ~~~~~~~~~~~~~~~~~~~~~~~~~&  \\
  \frac{\beta^2}{4} \frac{1}{L^{2d}} \sum_{{\bm k},{\bm k}'}  \sum_{i,j\in A} e^{i ( {\bm k}-{\bm k}'  ) \cdot ( {\bm r}_i-{\bm r}_j  ) } & 
 (\epsilon_{\bm k} - \epsilon_{\bm k'})^2 + {\cal O}(\beta^3)~.  \nonumber
\end{eqnarray}
 Comparing again with Eq.~\eqref{e.QVhighT}, one can conclude that, at high temperatures
 \begin{equation}
 \langle \delta^2 \hat{N}_A \rangle_Q  =   \frac{1}{3}~\frac{F_Q(\hat{N}_A,\hat{\rho})}{4}  +  {\cal O}(\beta^3)~.
 \end{equation}
 
 \subsection{Discussion}
 
Hence, as anticipated in the main text, the quantum fluctuations captured by the quantum variance, the WY skew information or the QFI display the same high-temperature behavior up to a global prefactor. This leads to a coherent picture for bipartite quantum fluctuations of free fermions. While the calculation of the quantum variance is easily extended to arbitrary many-body systems which can be treated with state-of-the-art numerics, the same is generally not true for the WY skew information nor the QFI -- although, unlike the QFI, the WY skew information lends itself to path-integral Monte Carlo approaches probing imaginary-time correlation functions. On the experimental side, the WY skew information, being an imaginary-time correlation function, is not accessible to experiments as such. As for the QFI, Ref.~\cite{Haukeetal2015} has recently shown that it is potentially accessible to experiments when cast as a frequency integral involving the dynamic susceptibility; in this respect, the quantum variance has the advantage of being expressed solely in terms of static correlations and response functions. 

\begin{center}
  \begin{figure}[h!]
  \includegraphics[width=6cm]{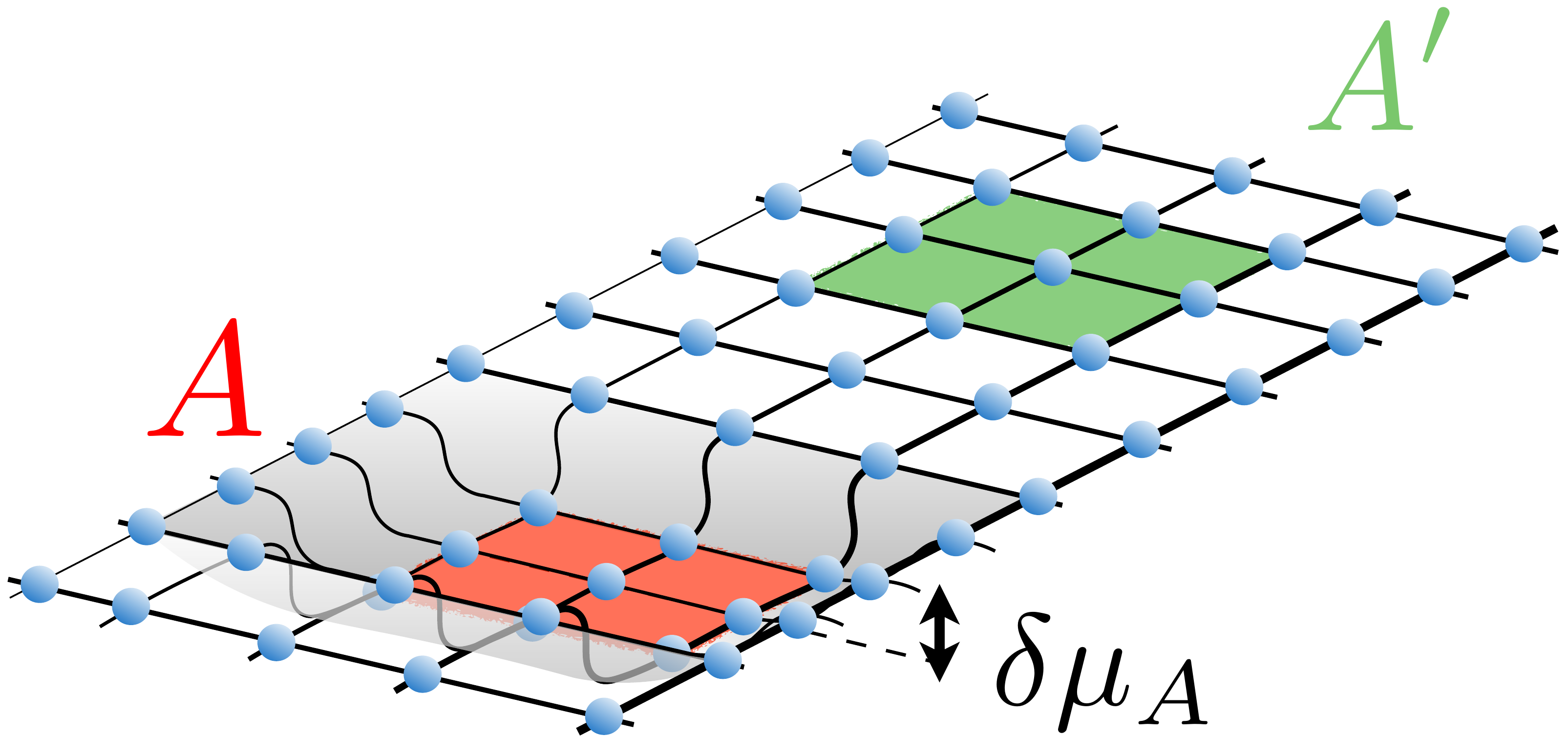}
  \caption{{\bf Quantum-gas microscope setup to measure the quantum variance.}  Here we sketch a possible scheme to measure the quantum variance of the local particle number in region $A$ by adding a box-like potential to a two-dimensional optical lattice. This potential induces a local increase in the chemical potential, allowing one to probe the response function as the particle-number difference between region $A'$ and $A$; supplementing this measurement with the one of particle-number variance in region $A$ gives access to the quantum variance.}
\label{f.microscope}
\end{figure}
\end{center}

\section{Measurement of bipartite quantum variance with quantum-gas microscopes}
\label{a.exp}

A concrete proposal to measure the quantum variance of bipartite particle-number fluctuations in the context of ultra-cold quantum gases is illustrated in Fig.~\ref{f.microscope}. Recent progress in quantum gas microscopes allows one to trap atoms in single layers of an optical lattice with superimposed traps of arbitrary geometries \cite{Bakretal2009}  and to measure single-site particle-number occupations \cite{Preissetal2015}. As already discussed in App.~\ref{a.pathintegral}, to access the quantum variance of the particle number $N_A$ in the subsystem $A$ one needs to measure the total variance of fluctuations $\langle \delta^2 N_A \rangle$, as well as the response function 
\begin{equation}
\chi_{N_A} \approx  \frac{\langle N_A\rangle (\mu_A+\delta\mu_A) - \langle N_A \rangle (\mu_A)}{\delta \mu_A}     
\end{equation}
where $\mu_A$ is the local chemical potential in region $A$, coupling to the particle number $N_A$. The two quantities  $\langle \delta^2 N_A \rangle$ and $\chi_{N_A}$ need to be measured in the same conditions of temperature and (offset) chemical potential. A way to achieve this in cold-atom experiments is to use a ``multiplexing" setup as in Fig.~\ref{f.microscope}, in which one single trap geometry allows one to measure both quantities at once. Indeed monitoring fluctuations of $N_A$ in region $A$ allows one to extract  $\langle \delta^2 N_A \rangle$ the total variance; on the other hand, a box-like potential superimposed to the optical lattice creates a local increase in the chemical potential, giving access to  the response function as $(\langle N_{A'} \rangle - \langle N_A \rangle)/\delta\mu_A$. If the regions $A$ and $A'$ are built symmetrically around the (global) trap center, and if thermal equilibrium is established across the system, one is ensured that the two quantities are measured in the same thermodynamic conditions of temperature and offset chemical potential. 

One may worry that in cold-atom experiments the total particle number has wide shot-to-shot fluctuations going well beyond a grand-canonical description, and that this may alter the estimate of the quantum variance, adding spurious contributions coming from experimental systematics. On the other hand, as discussed in the main text and in sections App.~\ref{a.coh} and \ref{a.pathintegral}, all incoherent fluctuations (either stemming from the grand-canonical ensemble or from other sources) are systematically subtracted away in the quantum variance, if one is able to realize experimentally the deformation of the density matrix as in \eqref{e.thermalrho_def}. We argue that this is indeed the case when the total particle number obeys an arbitrary statistics, namely the case in which the density matrix takes the general form
\begin{equation}
\hat{\rho}  = \frac{1}{\cal Z} \sum_N p_{\rm exp}(N) ~\hat{\cal P}_N e^{-\beta \hat{\cal H}}  \hat{\cal P}_N
\end{equation}
where $p_{\rm exp}(N)$ is the experimental particle-number statistics, accounting for systematic shot-to-shot fluctuations, and  $\hat{\cal P}_N$ is the projector onto the Fock subspace with $N$ particles. The deformation of the Hamiltonian implied in Fig.~\ref{f.microscope} leads to the desired deformation of the density matrix; hence the quantum variance (and its peculiar size and temperature scaling) can be experimentally measured even without postselection of the measurement shots according to the total particle number, with the obvious caveat that one is not measuring properties of the grand-canonical ensemble but the ones of the artificial ensemble realized experimentally. 

A similar setup, and similar considerations, can be applied to measure the quantum variance of the staggered particle number. In that case, one needs to shine a weak superlattice potential with twice the lattice spacing of the primary potential over the region $A'$.

\section{Quantum variance as multiparticle entanglement witness}
\label{a.enta}

Let us consider a system of $N$ qubits, with collective spin operators $\hat{\bm J} = \sum_{i=1}^{N} \hat{\bm S}_i$. A pure state $|\psi\rangle$ is said to be $k$-producible \cite{SeevinkU2001,Tothetal2005,Chen2005} if it can be written as
 \begin{equation}
 |\psi_{k{\rm -prod}} \rangle = \otimes{l=1}^{M} |\psi_{N_l}\rangle 
 \end{equation} 
 where $|\psi_{N_l}\rangle$ is an (entangled) state of a block of $N_l \leq k$ spins, with the constraint that $\sum_l N_l = N$. A mixed state is then said to be $k$-producible if it is an incoherent superposition of $k_s$-producible states with $k_s \leq k$
 \begin{equation}
 \hat{\rho}_{k{\rm-prod}} = \sum_s p_s |\psi_{k_s{\rm-prod}} \rangle \langle \psi_{k_s{\rm-prod}} |~.
 \end{equation} 
 Using Eq.~\eqref{e.qv_qfi} and the results of Refs.~\cite{Hyllusetal2012,Toth2012}, one can prove that for $k$-producible states the quantum variance of the collective spin components $\hat{J}^{\alpha}$, and the QFI associated to transformation generated by the $\hat{J}^{\alpha}$, satisfy the inequality:
 \begin{equation}
 \langle \delta^2 J^{\alpha} \rangle_Q \leq F_Q(J^{\alpha};\rho_{k{\rm-prod}})/4 \leq  nk^2  + (N-nk)^2 
 \label{e.kprod}
 \end{equation}
 where  $n = [N/k]$ is the integer part of $N/k$. In fact the exact same bound as in the last inequality of Eq.~\eqref{e.kprod} holds for the WY skew information \cite{Chen2005}, and again it carries over to the quantum variance thanks to the inequality in Eq.~\eqref{e.qv_WY}. {The inequality of Eq.~\eqref{e.kprod} can be readily generalized to more general degrees of freedom than qubits, namely to collective operators $\hat{C}  = \sum_i \hat{c}_i$ where $c_i$ is an operator with a bounded spectrum contained in the interval $[c_{\rm min}, c_{\rm max}]$. In that case the inequality takes the form \cite{PezzeS2014}
 \begin{eqnarray}
 4 \langle \delta^2 \hat{C} \rangle_Q & \leq &  F_Q(\hat C;\rho_{k{\rm-prod}}) \nonumber \\
 & \leq &  (c_{\rm max} - c_{\rm min})^2 [ nk^2  + (N-nk)^2 ]~. 
  \label{e.kprod_gen}
 \end{eqnarray}}
 
 Hence, similarly to what was already found for the WY skew information \cite{Chen2005} and the QFI \cite{Hyllusetal2012,Toth2012,PezzeS2014}, a violation of the inequalities in Eqs.~\eqref{e.kprod} or \eqref{e.kprod_gen} or  for the quantum variance is a strong indication of the existence of multiparticle entanglement among a least $(k+1)$ degrees of freedom. The condition of violation is actually very strong, as the bound of the last inequality in is rather loose for thermal states. { Indeed the bound of Eq.~\eqref{e.kprod} is valid for $k$-producible pure states and mixed states alike, but, given that all the quantities in question (WY skew information, QFI, and quantum variance) are expected to decrease under thermal mixing, the bound is much looser for thermal states, and all the more so the higher the temperature. }

\bibliography{qv}

\end{document}